\documentclass[twocolumn]{aastex62}

\usepackage{bm}
\usepackage{url}
\usepackage{amsmath}
\usepackage{color}
\usepackage{graphicx}

\received{August 23, 2018}
\revised{November 18, 2018}
\accepted{November 24, 2018}
\submitjournal{ApJ}

\shorttitle{dust-pileup}
\shortauthors{Ueda et al.}

\begin{document}

\title{
DUST-PILEUP AT THE DEAD-ZONE INNER EDGE AND IMPLICATIONS FOR THE DISK SHADOW
}

\correspondingauthor{Takahiro Ueda}
\email{t.ueda@geo.titech.ac.jp}

\author{Takahiro Ueda}
\affil{Department of Earth and Planetary Sciences, Tokyo Institute of Technology, Meguro, Tokyo, 152-8551, Japan}

\author{Mario Flock}
\affiliation{Jet Propulsion Laboratory, California Institute of Technology, Pasadena, California 91109, USA}
\affiliation{Max Planck Institute for Astronomy, K\"{o}nigstuhl 17, D-69117 Heidelberg, Germany}

\author{Satoshi Okuzumi}
\affiliation{Department of Earth and Planetary Sciences, Tokyo Institute of Technology, Meguro, Tokyo, 152-8551, Japan}



\begin{abstract}
We perform simulations of the dust and gas disk evolution to investigate the observational features of a dust-pileup at the dead-zone inner edge.
We show that the total mass of accumulated dust particles is sensitive to the turbulence strength in the dead zone, $\alpha_{\rm dead}$, because of the combined effect of turbulence-induced particle fragmentation (which suppresses particle radial drift) and turbulent diffusion.
For a typical critical fragmentation velocity of silicate dust particles of $1~{\rm m~s^{-1}}$, the stress to pressure ratio $\alpha_{\rm dead}$ needs to be lower than $3 \times 10^{-4}$ for dust trapping to operate.
The obtained dust distribution is postprocessed using the radiative transfer code RADMC-3D to simulate infrared scattered-light images of the inner part of protoplanetary disks with a dust pileup.
We find that a dust pileup at the dead-zone inner edge, if present, casts a shadow extending out to $\sim 10~{\rm au}$.
In the shadowed region the temperature significantly drops, which in some cases yields even multiple water snow lines.
We also find that even without a dust pileup at the dead-zone inner edge, the disk surface can become thermally unstable, and the excited waves can naturally produce shadows and ring-like structures in observed images. This mechanism might account for the ring-like structures seen in the scattered-light images of some disks, such as the TW Hya disk.
\end{abstract}



\section{Introduction} \label{sec:intro}
The inner region of protoplanetary disks is the birthplace of rocky planetesimals and planets.
One preferential site of rocky planetesimal formation is the inner edge of the so-called dead zone (e.g., \citealt{Kretke2009}).
The dead zone is the location where magneto-rotational instability (MRI, \citealt{Balbus1998}) is suppressed because of poor gas ionization \citep{Gammie1996}.
The dead zone is likely to have an inner edge where the gas temperature $T$ reaches $\sim$1000~K, above which thermal ionization of the gas is effective enough to activate MRI \citep{Gammie1996,Desch2015}.
Across the dead zone inner edge, the turbulent viscosity arising from MRI steeply decreases from inside out, resulting in a local maximum in the radial profile of the gas pressure (e.g., \citealt{Dzyurkevich2010}; \citealt{Flock2016}; \citealt{Flock2017}).
The pressure maximum traps solid particles \citep{Whipple1972, Adachi1976} and the local dust-to-gas mass ratio increases, leading potentially to the formation of rocky planetesimals via the streaming instability \citep{Youdin2005, Johansen2007, Carrera2015} or via the gravitational instability \citep{CMF1981}.

There have been many studies related to the dust-pileup at the dead-zone inner edge (e.g., \citealt{Brauer2008}; \citealt{Kretke2009}; \citealt{Dzyurkevich2010}; \citealt{Pinilla2016}).
\citet{Kretke2009} examined the accumulation and coagulation of solid particles around the dead-zone inner edge and found that the dust-pileup at the dead-zone inner edge leads to the efficient formation of gas giants.
However, \citet{Kretke2009} ignored the effect of the fragmentation of large silicate particles. 
Both laboratory experiments (e.g., \citealt{Blum2000}) and numerical simulations (e.g., \citealt{Wada2013}) have shown that silicate dust particles are easy to fragment with a typical collisional velocity in protoplanetary disks.
The fragmentation must regulate growth of solids and it significantly affect the behavior of dust motion in the gas.
There are also some studies on the dust-pileup at the dead-zone inner edge caused by the combination of the inner-hole due to disk wind and non-thermal ionization (e.g., \citealt{Pinilla2016}) or by the ice sublimation (e.g., \citealt{Brauer2008}).
In this work, we focus on the innermost dust concentration zone, which is of most importance also in  the inside-out planet formation framework (e.g., \citealt{Chatterjee2014}).

Even for the observational aspects, the dead-zone inner edge would be a interesting subject.
For example, VLTI/MATISSE instrument will allow us to image the inner region of protoplanetary disks at mid-infrared wavelengths with high angular resolution ($\sim 5~{\rm mas}$ in L-band, e.g., \citealt{Lopez2014}), which will provide us opportunities to directly compare the theoretical models with the observations of the dead-zone inner edge.
Especially nearby Herbig stars are ideally suited to study this region due to high luminosity and the position of the inner dust rim (\citealt{Dullemond2010} for review).
However, previous studies on the dust-pileup at the dead-zone inner edge have not focused on the comparison with the observations. 

This work mainly consists of two parts.
First, we investigate the properties of the dust-pileup at the dead-zone inner edge for a broad range of critical fragmentation velocity of silicate dust particles and strength of the turbulence.
Second, we examine the observational signatures of the dust-pileup at the dead-zone inner edge by performing the radiative transfer simulations.
In Section \ref{sec:pileup}, we introduce the model of dust growth simulations and provide results of the simulations.
The setup for the radiative transfer simulations and its results are shown in Section \ref{sec:rad}.
In Section \ref{sec:instability}, we briefly mention the non-convergence found in the radiative transfer simulations probably caused by the so-called thermal wave instability.
The implications for the planet formation and for disk observations are given in Section \ref{sec:diss}. 
The summary is in Section \ref{sec:sum}.

\section{Dust-pileup at the dead-zone inner edge}\label{sec:pileup}
In this section, we introduce models of dust growth simulation and show how the properties of the dust-pileup depend on the critical fragmentation velocity of dust particles and the strength of the turbulence.
\subsection{Dust and gas evolution models}
We investigate the evolution of dust and gas disk around a Herbig-type star with stellar mass $M_{*}=2.5M_{\odot}$, radius $R_{*}=2.5R_{\odot}$ and effective temperature $T_{*}=10000~{\rm K}$. 
The resultant stellar luminosity is $56L_{\odot}$.
The size evolution of dust particles is also simultaneously calculated and we employ simplified dust coagulation equation in which the dust size distribution is characterized by the single representative mass $m_{\rm p}(r)$.
\subsubsection{Surface density evolution}
We follow the time evolution of the gas and dust surface densities $\Sigma_{\rm g}$ and $\Sigma_{\rm d}$ by calculating one-dimensional continuity equation of dust
\begin{eqnarray}
\frac{ \partial \Sigma_{\rm d} }{\partial t} + \frac{1}{r} \frac{\partial}{\partial r} \left\{ rv_{\rm r,d}\Sigma_{\rm d} -\frac{\nu}{1+{\rm St}^{2}}r\Sigma_{\rm g}\frac{\partial Z}{\partial r}\right\}= 0,
\end{eqnarray}
and that of gas
\begin{eqnarray}
\frac{ \partial \Sigma_{\rm g} }{\partial t} + \frac{1}{r} \frac{\partial}{\partial r} (rv_{\rm r,g}\Sigma_{\rm g}) = 0,
\end{eqnarray}
where $r$ is a midplane distance from the central star, $\nu$ is the turbulent viscosity, 
St is the dimensionless stopping time of the dust particles, $Z$ is the dust-to-gas surface density ratio, $v_{\rm r,d}$ and $v_{\rm r,g}$ are the radial velocities of dust and gas, respectively.
The radial velocities of dust and gas are respectively written as \citep{Kretke2009, Kanagawa2017}
\begin{eqnarray}
v_{\rm r,d}=-\frac{\rm St}{{\rm St^{2}}+(1+Z)^{2}}2\eta v_{\rm K}+ \frac{1+Z}{{\rm St^{2}}+(1+Z)^{2}} v_{\rm vis}
\label{eq:dustvel}
\end{eqnarray}
and
\begin{eqnarray}
v_{\rm r,g}=\frac{{\rm St}Z}{{\rm St^{2}}+(1+Z)^{2}}2\eta v_{\rm K}+ \left\{ 1-\frac{(1+Z)Z}{{\rm St^{2}}+(1+Z)^{2}} \right\} v_{\rm vis},
\label{eq:gasvel}
\end{eqnarray}
where $v_{\rm K}$ is the Keplerian velocity and $v_{\rm vis}$ is the radial velocity of the gas due to viscous diffusion written as \begin{eqnarray}
v_{\rm vis}=-\frac{3\nu}{r} \frac{\partial \ln \left( r^{1/2}\nu \Sigma_{\rm g} \right) }{\partial \ln r}.
\label{eq:vvis}
\end{eqnarray}
Equations \eqref{eq:dustvel} and \eqref{eq:gasvel} fully include the effect of dust backreaction on the gas disk.
When the disk turbulence is weak and the dust-to-gas mass ratio is high enough, the backreaction leads to the outward motion of the gas, which facilitates efficient piling-up of dust particles \citep{Gonzalez2017, Kanagawa2017}. 
The variable $\eta$ characterizes the sub-Keplerian motion of the gas disk and is written as
\begin{eqnarray}
\eta=-\frac{1}{2} \left( \frac{c_{s}}{v_{\rm K}} \right)^{2} \frac{\partial \ln p}{\partial \ln r},
\label{eq:eta}
\end{eqnarray}
where $c_{\rm s}=\sqrt{k_{\rm B}T/m_{\rm g}}$ is the sound speed of disk gas at the midplane and $p=\rho_{\rm g}c_{\rm s}^{2}$ is pressure of the gas with $k_{\rm B}$ and $m_{\rm g}$ being the Boltzmann constant and mean molecular mass (taken to be 2.4 amu), respectively.

The motion of dust in disks is characterized by the Stokes number, dimensionless stopping time, defined as
\begin{eqnarray}
{\rm St} \equiv \Omega_{\rm K}t_{\rm s},
\end{eqnarray}
where $\Omega_{\rm K}$ is the Keplerian frequency and $t_{\rm s}$ is the stopping time of the dust particles.
The stopping time is related to the dust radius $a$ as
\begin{eqnarray}
t_{\rm s}=
\begin{cases}
{\displaystyle \frac{\rho_{\rm int}a}{\rho_{\rm g}v_{\rm th}}} ,  & \ {\displaystyle a < \frac{9\lambda_{\rm mfp}}{4}},\\
{\displaystyle \frac{\rho_{\rm int}a}{\rho_{\rm g}v_{\rm th}} \frac{4a}{9\lambda_{\rm mfp}}}, & \ {\displaystyle a > \frac{9\lambda_{\rm mfp}}{4}},
\label{eq:ts}
\end{cases}
\end{eqnarray}
where $\rho_{\rm int}$ is the dust internal density, $\rho_{\rm g}$ is the midplane gas density, $v_{\rm th}=\sqrt{8/\pi}c_{\rm s}$ is the thermal velocity of the gas and $\lambda_{\rm mfp}$ is the mean free path of gas molecules. 
The midplane gas density is given by $\rho_{\rm g}=\Sigma_{\rm g}/\sqrt{2\pi}h_{\rm g}$, where $h_{\rm g}=c_{\rm s}/\Omega_{\rm K}$ is the gas scale height.
The mean free path of gas molecules is related to the midplane gas density as $\lambda_{\rm mfp}=m_{\rm g}/(\sigma_{\rm mol}\rho_{\rm g})$, where $\sigma_{\rm mol}=2\times10^{-15}~{\rm cm^{2}}$ is the molecular collisional cross section. 
From these, the Stokes number can be rewritten as
\begin{eqnarray}
{\rm St} = \frac{\pi}{2} \frac{\rho_{\rm int}a}{\Sigma_{\rm g}} \max \left( 1, \frac{4a}{9\lambda_{\rm mfp}} \right).
\end{eqnarray}
For simplicity, the dust internal density is set to be $3.0~{\rm g~cm^{-3}}$ for the region where the icy component is evaporated (i.e., $T>160~{\rm K}$) and $1.4~{\rm g~cm^{-3}}$ elsewhere.
The initial gas surface density is calculated assuming the radially constant mass accretion rate of $10^{-8}M_{\odot}~{\rm yr^{-1}}$ and the initial dust surface density is set to be $0.01\Sigma_{\rm g}$.

\subsubsection{Dust-size evolution}
For the evolution of dust particles, we calculate the evolution of a representative mass of dust particles, $m_{\rm p}$, for each radial grid using the single-size approximation \citep{Sato2016}:
\begin{eqnarray}
\frac{ \partial m_{\rm p} }{\partial t} + v_{\rm r,d}\frac{\partial m_{\rm p}}{\partial r} = \epsilon_{\rm grow}\frac{2\sqrt{\pi}a^{2}\Delta v_{\rm pp}}{h_{\rm d}} \Sigma_{\rm d},
\end{eqnarray}
where $h_{\rm d}$ is the scale-height of the dust disk.
The dust scale height is assumed to be a mixing-settling equilibrium \citep{Dubrulle1995, Youdin2007}
\begin{eqnarray}
h_{\rm d}=h_{\rm g} \left( 1+ \frac{\rm St}{\alpha}\frac{1+2{\rm St}}{1+{\rm St}}\right)^{-1/2},
\label{eq:dustheight}
\end{eqnarray}
where $\alpha$ is the stress to pressure ratio \citep{shakura}.
The coefficient $\epsilon_{\rm grow}$ is the sticking efficiency for a single collision, which we model as \citep{OH2012, Okuzumi2016}
\begin{eqnarray}
\epsilon_{\rm grow}= \min \left\{ 1,-\frac{\ln (\Delta v_{\rm pp}/v_{\rm frag})}{\ln 5} \right\},
\end{eqnarray}
where $\Delta v_{\rm pp}$ is the relative velocity between colliding particles and $v_{\rm frag}$ is the critical fragmentation velocity determined by the mechanical properties of the particles.
If $\Delta v_{\rm pp}>v_{\rm frag}$, $\epsilon_{\rm grow}$ is negative, meaning that the single collision results into the fragmentation of colliding particles.
Laboratory experiments (e.g., \citealt{Blum2000}) and numerical simulations (e.g., \citealt{Wada2013}) have shown that the typical value of $v_{\rm frag}$ for silicate aggregates is in the range 1--10 ${\rm m\,s^{-1}}$, with the exact value depending on the size of the grains constituting the aggregates. 
We change the value from $0.1$ to $10~{\rm m~s^{-1}}$ and investigate how it affects the dust evolution around the dead-zone inner edge.
The fragmentation velocity of icy dust particles is known to be higher than that for silicate particles \citep{Wada2013, GB2015}. 
Because the region beyond the snow line is not our main focus, we simply adopt $v_{\rm frag}=30~{\rm m~s^{-1}}$ for icy particles.
With this value of $v_{\rm frag}$, icy particles do not experience catastrophic disruption.

For the components of the relative velocity between colliding particles, we consider the relative velocity due to brownian motion of dust particles $\Delta v_{\rm B}$, disk turbulence $\Delta v_{\rm t}$, azimuthal velocity $\Delta v_{\rm \phi}$, settling velocity $\Delta v_{\rm z}$ and radial drift velocity $\Delta v_{\rm r}$.
Therefore, $\Delta v_{\rm pp}$ is written as
\begin{eqnarray}
\Delta v_{\rm pp} =\sqrt{\Delta v_{\rm B}^{2} + \Delta v_{\rm r}^{2} + \Delta v_{\rm \phi}^{2} + \Delta v_{\rm z}^{2} + \Delta v_{\rm t}^{2}}.
\label{eq:relv}
\end{eqnarray}
Around the dead-zone inner edge, due to the high temperature, the relative velocity is dominated by the velocity originating from the disk turbulence \citep{Ormel2007}
\begin{eqnarray}
\Delta v_{\rm t}=
\begin{cases}
\sqrt{\alpha}c_{\rm s}{\displaystyle \sqrt{\frac{1}{1+{\rm St}}+\frac{1}{1+{\rm \epsilon St}}} },&\ {\rm St}>1,\\
\sqrt{3\alpha \rm St}c_{\rm s},&\ t_{\eta}\Omega_{\rm K}< {\rm St} \leq 1,\\
\sqrt{\alpha}c_{\rm s} {\rm Re}_{t}^{1/4}(1-\epsilon){\rm St},&\ {\rm St} \leq t_{\eta}\Omega_{\rm K},
\label{eq:relvt}
\end{cases}
\end{eqnarray}
where ${\rm Re}_{\rm t}=2\nu/v_{\rm th}\lambda_{\rm mfp}$ is the turbulent Reynolds number, $t_{\eta}={\rm Re}_{\rm t}^{-1/2}/\Omega_{\rm K}$ is the turnover timescale of the smallest eddies and $\epsilon=0.5$ represents the ratio of the Stokes number of colliding particles.
\citet{Sato2016} found the best fit value of $\epsilon=0.5$ for this model which means that the dominant collisions are represented by collisions of grains with a size ratio of two.
For full details of the velocity components, we refer the readers to \citet{Sato2016}.

We also consider the sublimation of silicate and icy components. 
We assume that almost all silicate particles in a grid sublimate if the temperature in the grid is higher than $1350~{\rm K}$, although in order to stabilize calculations we keep a very tiny amount of dust ($\Sigma_{\rm d}=10^{-10}\Sigma_{\rm g}$) there.
Icy particles are assumed to sublimate on the water snow line, which we define as the radial position where the temperature reaches $160~{\rm K}$. We include this effect by reducing the inward solid mass flux across the snow line by 50\%, which is the assumed ice fraction of the solid particles in the outer disk.
For simplicity, we ignore the re-condensation of silicate and water ice.
The initial dust radius is assumed to be $0.1~{\rm \mu m}$ for the entire region of the disk.
 
\subsubsection{Radial temperature profile}
We focus on passive protoplanetary disks, where the radiation from the central star dominates.
The temperature profile is simply assumed to be the temperature profile of an optically thin disk,
\begin{eqnarray}
T=\epsilon_{\rm emit}^{-1/4}\left(\frac{R_{*}}{2r}\right)^{1/2}T_{*},
\label{eq:temp}
\end{eqnarray}
where $\epsilon_{\rm emit}=1/3$ is a ratio between the emission and absorption efficiencies of dust particles.
Although this temperature profile is not valid in the optically thick region, we use Equation \eqref{eq:temp} because our main focus is on the inner-most region where the temperature can be well described by an optically thin passive disk model \citep{Ueda2017, Flock2017}. 
The actual temperature profile would depend on the dust distribution and evolve with time, but in the dust-growth simulations, we do not consider the evolution of the temperature structure. 
As mentioned later, in the radiative transfer simulations, the vertical structure is iteratively calculated using the temperature structure obtained from the previous simulation to obtain the vertically consistent model.
For simplicity, we ignore the effect of accretion heating as the effect remains low for Herbig stars with the given mass accretion rate \citep{Flock2017}.

\subsubsection{Turbulence}
For the turbulent viscosity, we use the $\alpha$-prescription \citep{shakura}
\begin{eqnarray}
\nu=\alpha c_{\rm s}h_{\rm g}.
\label{eq:viscosity}
\end{eqnarray}
We set $\alpha$ as a function of the midplane temperature as \citep{Flock2016}
\begin{eqnarray}
\alpha=\frac{(\alpha_{\rm MRI}-\alpha_{\rm dead})}{2} \left[ 1-\tanh{\displaystyle \left(\frac{T_{\rm MRI}-T}{50~\rm K}\right)} \right]+\alpha_{\rm dead},
\label{eq:alpha}
\end{eqnarray}
where $T_{\rm MRI}=1000~{\rm K}$ is a critical temperature to activate MRI \citep{Desch2015}, $\alpha_{\rm dead}$ is the value of $\alpha$ in the MRI-inactive region and $\alpha_{\rm MRI}$ is that in the MRI-active region.
This description leads to $\alpha \approx \alpha_{\rm MRI}$ for $T \gg T_{\rm MRI}$ and $\alpha \approx \alpha_{\rm dead}$ for $T \ll T_{\rm MRI}$.
The value of $\alpha_{\rm dead}$ is quite uncertain, so we change the value from $10^{-4}$ to $10^{-2}$ and set $\alpha_{\rm MRI}$ to be 10 times larger than $\alpha_{\rm dead}$.

\subsubsection{Planetesimal formation}
If the dust-to-gas mass ratio is sufficiently high, part of dust particles would potentially be converted into plantesimals via the streaming instability (e.g., \citealt{Youdin2005}).
To take this effect into account, in some simulations, if the midplane dust density is higher than the gas density, we convert part of the dust surface density into planetesimals following the approach of \citet{DAM2016}:
\begin{eqnarray}
\frac{\partial\Sigma_{\rm plts}}{\partial t} =\zeta\frac{\Sigma_{\rm d}}{T_{\rm K}}, \quad \frac{\partial\Sigma_{\rm d}}{\partial t} = - \frac{\partial\Sigma_{\rm plts}}{\partial t},
\label{eq:plts}
\end{eqnarray}
where $\Sigma_{\rm plts}$ is the planetesimal surface density, $T_{\rm K}$ is the orbital period and $\zeta=10^{-4}$ is the planetesimal formation efficiency.
While \citet{DAM2016} assumed that only particles with ${\rm St}>10^{-2}$ are converted into planetesimals, we do not set such limitation on the size of dust particles because recent numerical simulations suggest that the streaming instability would occur even for smaller particles (${\rm St}<10^{-2}$) \citep{YJC2017}.

\subsection{Results of simulations of dust and gas disk evolution}
\label{sec:dustpileup}
\begin{figure*}[ht]
\begin{center}
\includegraphics[scale=0.35]{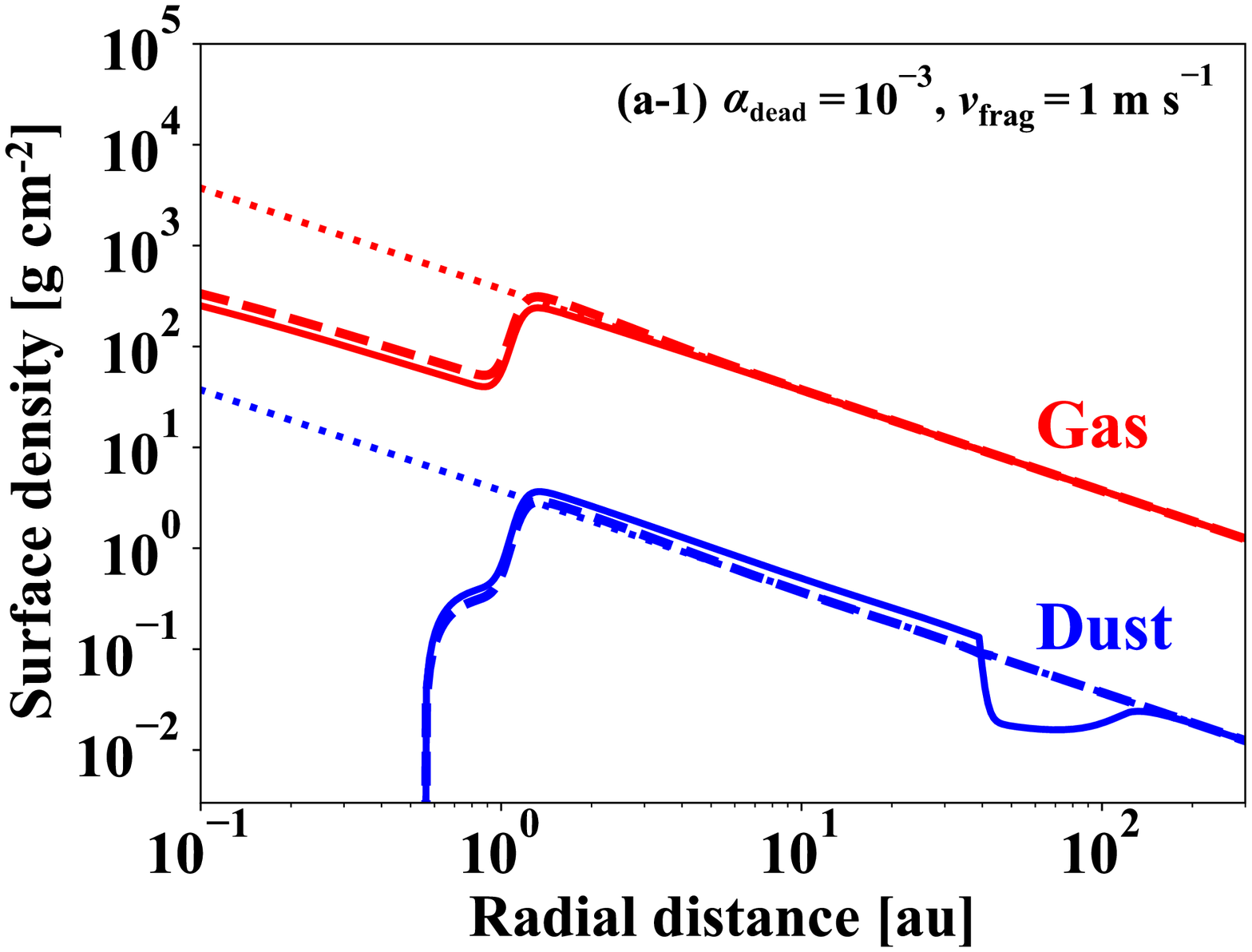}\includegraphics[scale=0.35]{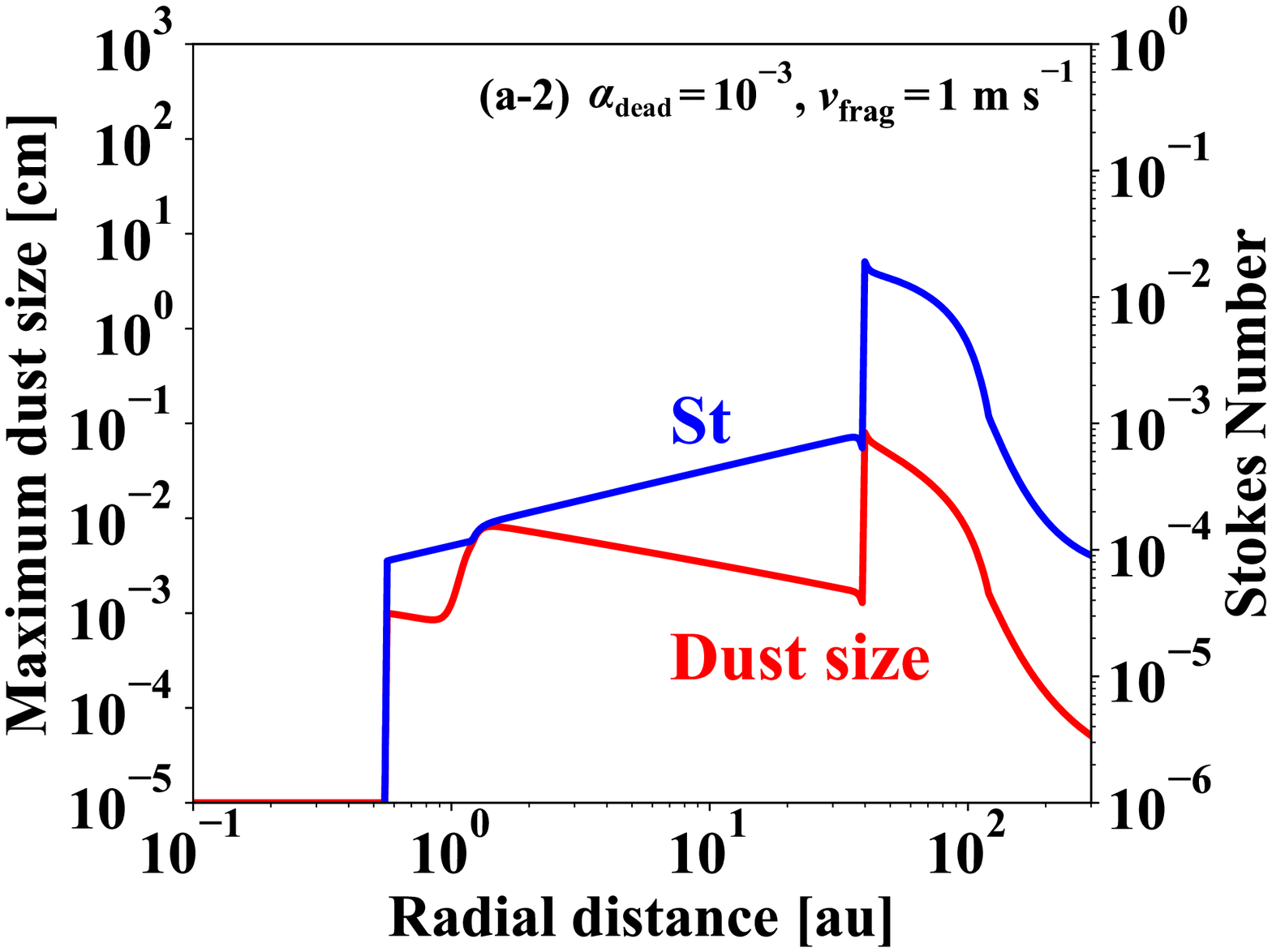}
\includegraphics[scale=0.35]{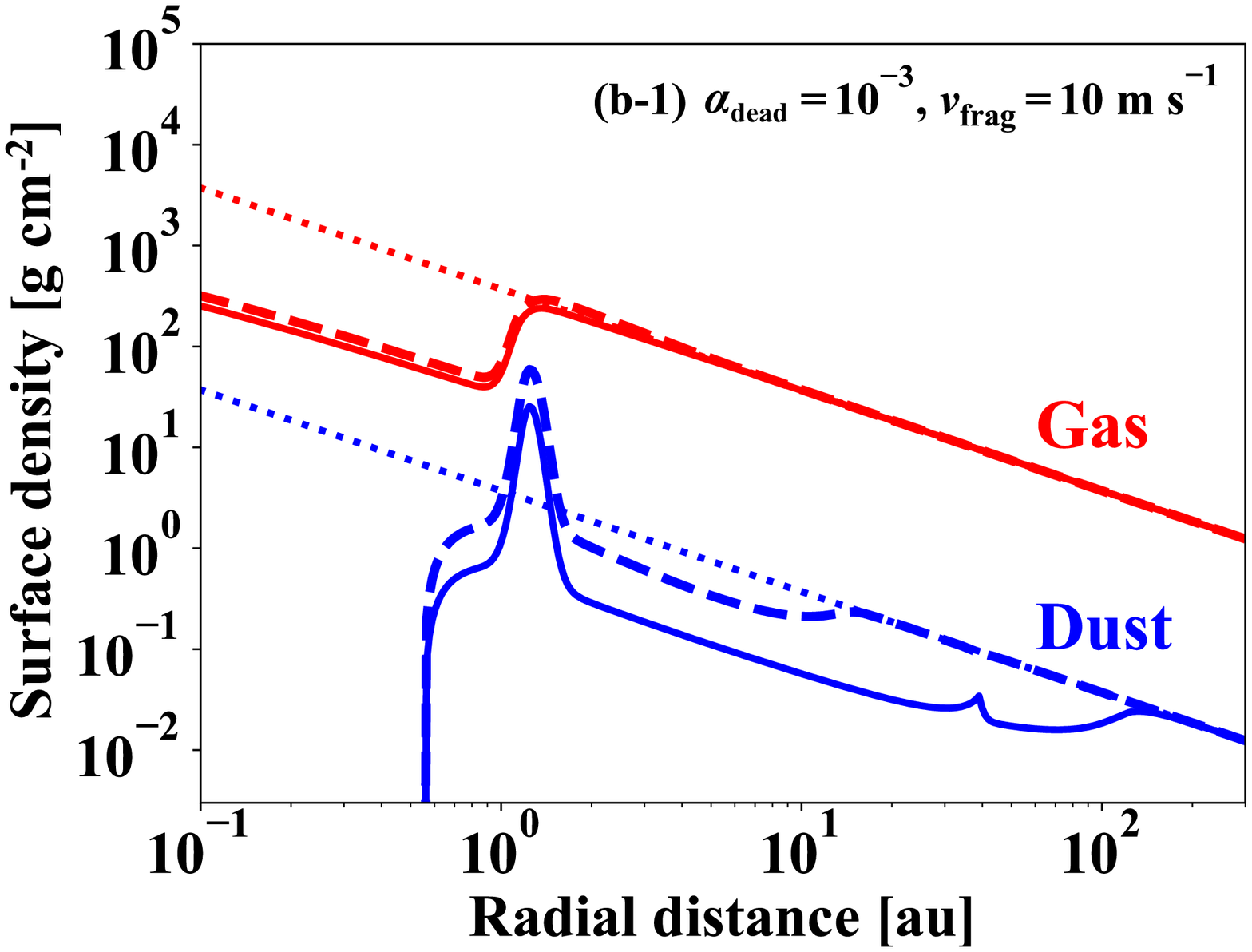}\includegraphics[scale=0.35]{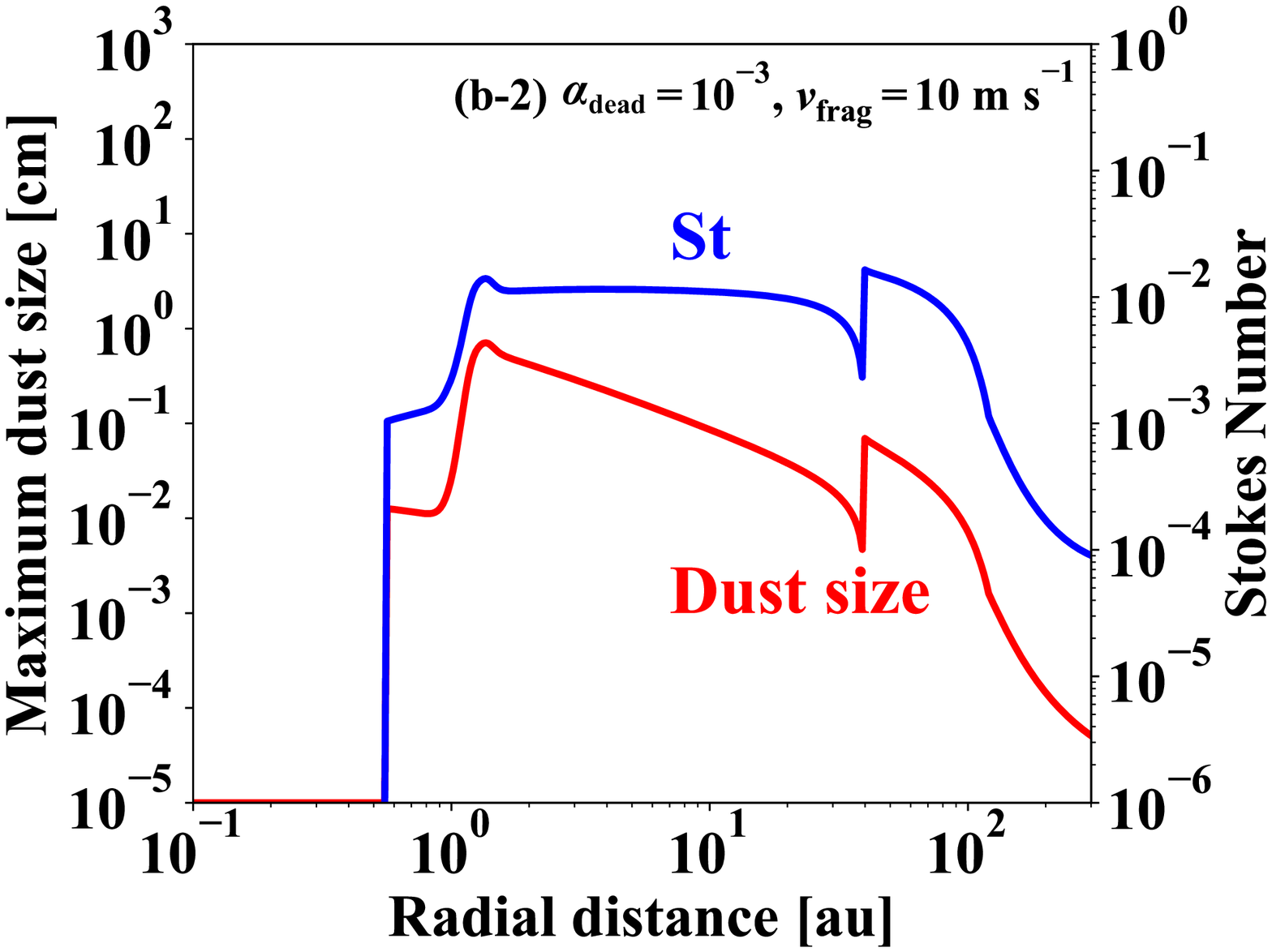}
\includegraphics[scale=0.35]{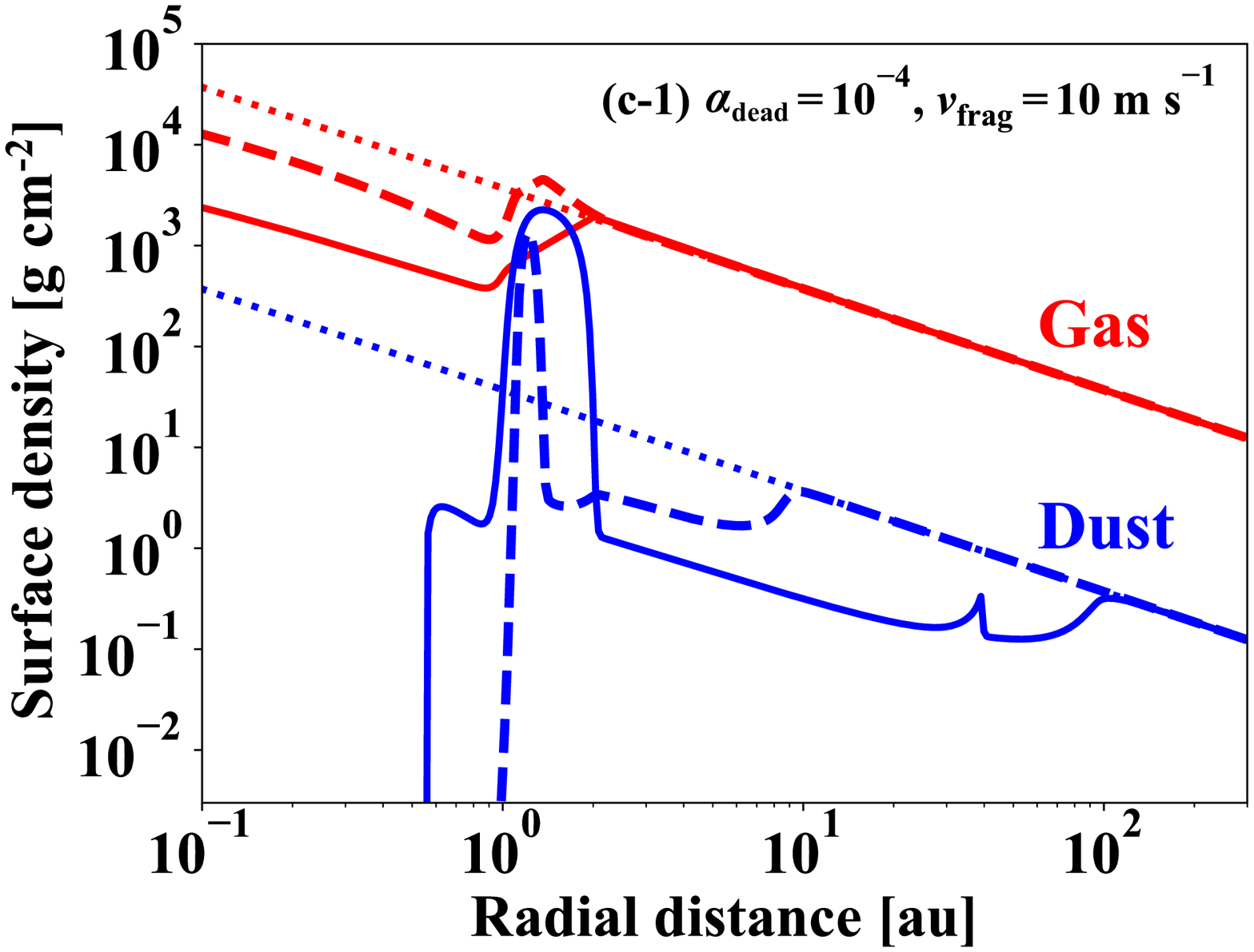}\includegraphics[scale=0.35]{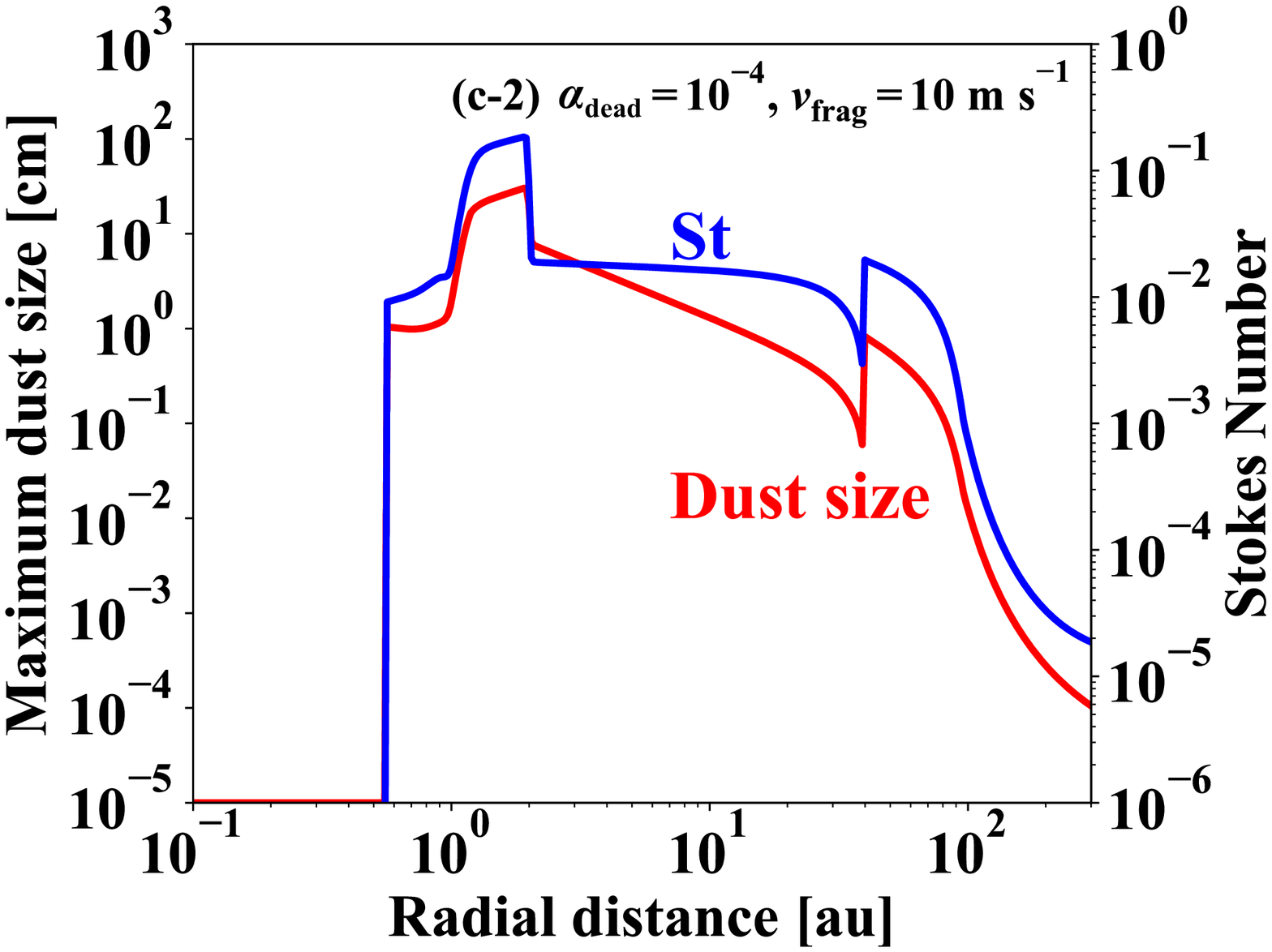}
\includegraphics[scale=0.35]{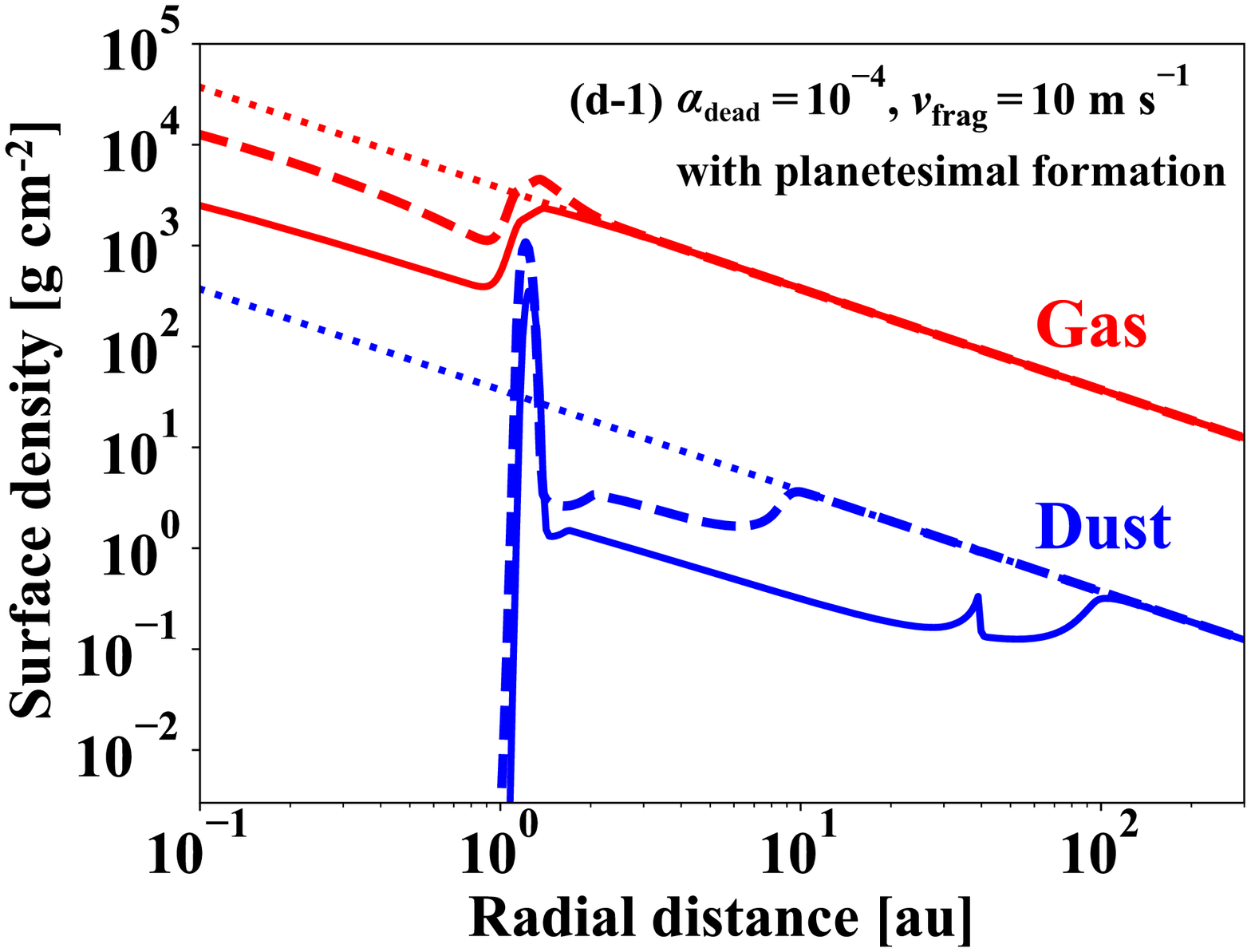}\includegraphics[scale=0.35]{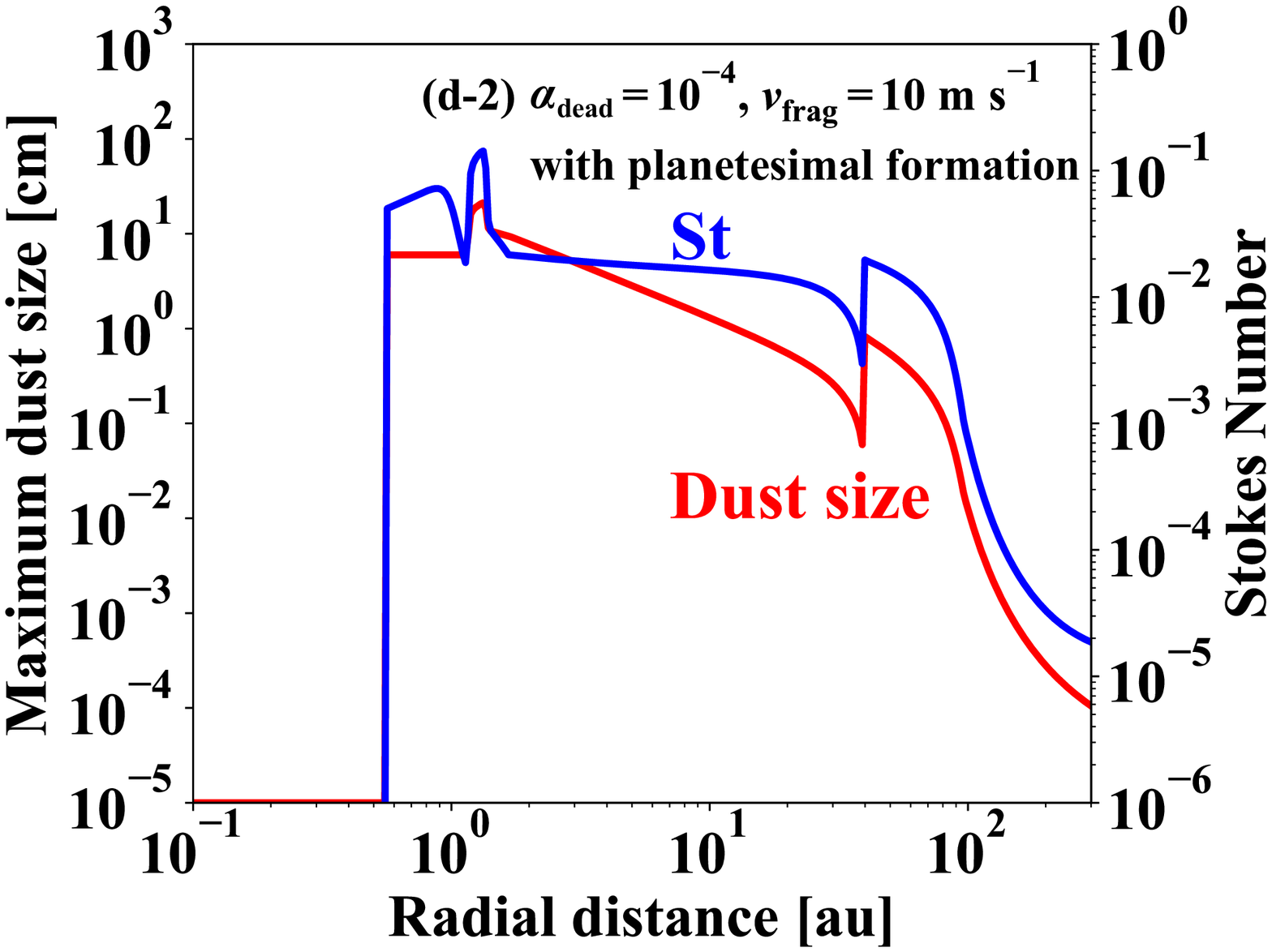}
\caption{
Dust and gas surface densities at $t=0~{\rm yr}$ (dotted), $1\times 10^{4}~{\rm yr}$ (dashed) and $3\times10^{5}~{\rm yr}$ (solid) (left-planels) and the maximum dust radius and the Stokes number at $t=3\times10^{5}~{\rm yr}$ (right-planels) for (a) $\alpha_{\rm dead}=10^{-3}$ and $v_{\rm frag}=1~{\rm m~s^{-1}}$, (b) $\alpha_{\rm dead}=10^{-3}$ and $v_{\rm frag}=10~{\rm m~s^{-1}}$, (c) $\alpha_{\rm dead}=10^{-4}$ and $v_{\rm frag}=10~{\rm m~s^{-1}}$,
(d) $\alpha_{\rm dead}=10^{-4}$ and $v_{\rm frag}=10~{\rm m~s^{-1}}$ with planetesimal formation (Equation \eqref{eq:plts}).
}
\label{fig:sigma}
\end{center}
\end{figure*}
Figure \ref{fig:sigma} shows the time evolution of the gas and dust disks with different values of $\alpha_{\rm dead}$ and $v_{\rm frag}$.
In our disk model for a Herbig type star, the inner rim of dust disk (location where silicate dust particles sublimate) and the dead-zone inner edge are located at $\sim0.5~{\rm au}$ and $\sim1~{\rm au}$, respectively.

Figure 1(a) shows the dust and gas surface density for three different time snapshots assuming $\alpha_{\rm dead}=10^{-3}$ and $v_{\rm frag}=1~{\rm m~s^{-1}}$. 
For this model, the inner disk ($<40$ au) has nearly reached a steady state. 
The dust particles remain small and no pileup of dust particles is seen at the dead-zone inner edge even though there is a local pressure maximum. This can be explained by the fact that, when the disk turbulence is strong and/or dust particles are poorly sticky (i.e., fragmentation velocity is small), dust particles keep their size to be so small that they cannot accumulate on the pressure maximum due to turbulent diffusion.

This picture of the disk evolution changes dramatically if one allows the grains to grow further. 
For $v_{\rm frag}=10~{\rm m~s^{-1}}$ (Figures 1(b), (c) and (d)), dust particles pile up at the dead-zone inner edge.
In Figure 1(a), the maximum Stokes number at the dead-zone inner edge is of the order of $10^{-4}$, which is $\sim$100--1000 times smaller than that in Figure 1(b), (c) and (d).
Even though there is a dust-pileup in Figure 1(b), the dust-to-gas mass ratio at the midplane remains above unity only for $5\times10^{4}$ years.
This is because the dust mass accretion rate decreases with time and the trapped dust particles gradually leaks out due to turbulent diffusion. 
For $\alpha_{\rm dead}=10^{-4}$ (Figure 1(c)) the dust-to-gas mass ratio at the midplane around the dead-zone inner edge exceeds unity for more than $5\times10^{5}$ years.

The evolution of dust and gas disk with the same parameter with Figure 1(c) but including the effect of planetesimal formation is shown in Figure 1(d).
In Figure 1(c), the width of the dust-concentrated region at $t=5\times 10^{5}~{\rm year}$ is $\sim 1~{\rm au}$, while it is narrower ($\sim 0.3~{\rm au}$) in Figure 1(d).
This is because in Figure 1(d), the mass transfer into planetesimals inhibits a strong dust accumulation.
If planetesimal formation occurs, dust particles trapped at the dead-zone inner boundary are converted into planetesimals before they leak out, resulting in the depletion of dust particles inside the dead-zone inner boundary.
Figure 1(d) shows large particles ($\sim 10~{\rm cm}$) between 0.6 and 1 au with a very small surface density ($< 10^{-3}~{\rm g~cm^{-3}}$).
This very tiny amount of large particles is leaking out of the dust-pileup. 
They keep their size large within the timescale we focus on because they hardly collide with each other owing to low dust-to-gas mass ratio.

In contrast, the outer disk region behind the snow line has not much evolved for the given time output. 
A unique feature is a drop of the size of dust particles when crossing the ice line (e.g., \citealt{Birnstiel2010, Okuzumi2016}).
Figure 1(a) and (b) show mm-sized grains at the snow line, while the grains can grow up to cm size at the snow line at 40 au  in Figure 1(c) and (d).

\begin{figure}[ht]
\begin{center}
\includegraphics[scale=0.5]{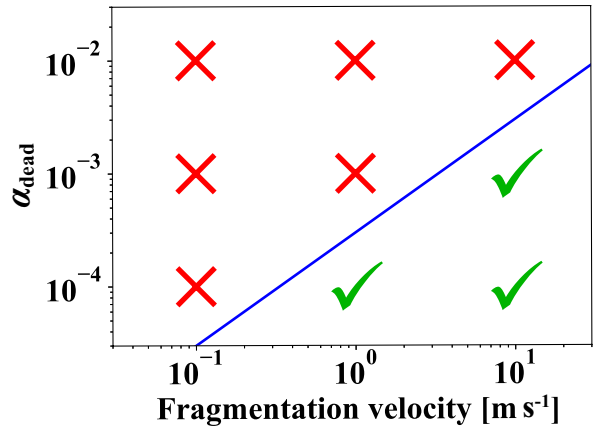}
\caption{Parameter dependence on the dust-pileup at the dead-zone inner edge. 
The green ticks and red crosses respectively denote the case where the disk has a dust-pileup and no dust-pileup at the dead-zone inner edge.
The blue solid line represents the criterion for the dust-pileup given by Equation \eqref{eq:pileup-condition2}.
}
\label{fig:pileup-condition}
\end{center}
\end{figure}
The results shown in Figure \ref{fig:sigma} imply that a strong dust-pileup at the dead-zone inner edge only occurs when the dust particles can grow large enough, either due to a reduced turbulent mixing or a higher fragmentation velocity. 
To confirm this over a wider parameter space, we summarize in Figure \ref{fig:pileup-condition} the outcome of all our simulations.
For $\alpha_{\rm dead}=10^{-2}$, dust particles cannot pile up on the dead-zone inner edge even if the critical fragmentation velocity of silicate dust particles is $10~{\rm m~s^{-1}}$.
As mentioned above, for the disk with $\alpha_{\rm dead}=10^{-3}$ and $v_{\rm frag}=10~{\rm m~s^{-1}}$, dust particles can pile up on the dead-zone inner edge, but the maximum dust-to-gas mass ratio at the midplane is merely $\sim3$ and is kept for less than $5\times10^{4}$ years. 
The total planetesimal mass is $1M_{\oplus}$ for the disk with $\alpha_{\rm dead}=10^{-3}$ and $v_{\rm frag}=10~{\rm m~s^{-1}}$, $374M_{\oplus}$ for the disk with $\alpha_{\rm dead}=10^{-4}$ and $v_{\rm frag}=1~{\rm m~s^{-1}}$, and $705M_{\oplus}$ for the disk with $\alpha_{\rm dead}=10^{-4}$ and $v_{\rm frag}=10~{\rm m~s^{-1}}$.

Here we derive the condition for the dust-pileup at the dead-zone inner edge as a function of $\alpha_{\rm dead}$ and $v_{\rm frag}$ by comparing the radial drift timescale of dust particles $\tau_{\rm drift}$ with the diffusion timescale $\tau_{\rm diff}$.
If the diffusion timescale is much larger than the radial drift timescale of dust particles , dust particles can pile up without a significant mass loss due to the diffusion (see also \citealt{Zhu2012}; \citealt{BDP2013}).
If we assume ${\rm St}\ll1$, which is basically valid in the whole disk (see Figure \ref{fig:sigma}), the condition can be written as
\begin{eqnarray}
\frac{\tau_{\rm vis}}{\tau_{\rm drift}}\sim\frac{r^{2}}{\nu} \frac{{\rm St}\eta v_{\rm K}}{r}=\frac{\rm St}{2\alpha} \left|-\frac{\partial \ln p}{\partial \ln r} \right| \gg 1.
\label{eq:pileup-condition1}
\end{eqnarray}
When we assume the absolute value of $\partial \ln p / \partial \ln r $ is an order of unity, Equation \eqref{eq:pileup-condition1} requires ${\rm St} \gg \alpha$.
Let us estimate the typical value of the Stokes number of dust particles at the dead-zone inner edge.
Because the size of dust particles in the inner region of disks is regulated by the collisional fragmentation due to turbulent motion, the size can be estimated by equating the relative velocity originating from the turbulence with the critical fragmentation velocity: $\sqrt{3\alpha {\rm St}}c_{\rm s}=v_{\rm frag}$.
Because the dust-pileup occurs within the MRI-suppressed region, i.e. $\alpha=\alpha_{\rm dead}$, the Stokes number of dust particles in the dust-concentrated region is written as
\begin{eqnarray}
{\rm St} \approx 10^{-4} \left( \frac{v_{\rm frag}}{1~{\rm m~s^{-1}}} \right)^{2} \left( \frac{\alpha_{\rm dead}}{10^{-3}} \right)^{-1},
\label{eq:typicalst}
\end{eqnarray}
here we use $T=T_{\rm MRI}=1000~{\rm K}$.
Equation \eqref{eq:typicalst} well explain the Stokes number shown in Figure \ref{fig:sigma}.
Therefore, the critical value of $\alpha_{\rm dead}$ for the dust-pileup at the dead-zone inner edge can be written as a function of $v_{\rm frag}$ as
\begin{eqnarray}
\alpha_{\rm dead} \approx 3\times 10^{-4} \left( \frac{v_{\rm frag}}{1~{\rm m~s^{-1}}} \right).
\label{eq:pileup-condition2}
\end{eqnarray}
If $\alpha_{\rm dead}$ is larger than this value, dust particles are unable to pile up on the dead-zone inner edge due to efficient diffusion.
In Figure \ref{fig:pileup-condition}, we indicate Equation \eqref{eq:pileup-condition2} by the blue solid line.
We find that Equation \eqref{eq:pileup-condition2} explains 
the outcomes of all our simulations.

\section{Radiative Transfer Simulations}\label{sec:rad}
The dust wall formed at the dead-zone inner boundary might block off the stellar irradiation and cast a shadow just behind it. 
In order to investigate this effect on the disk structure and its appearance, we perform radiative transfer calculations with the Monte Carlo radiative transfer code RADMC-3D \citep{RADMC}.

\subsection{Radiative Transfer Models}
We perform radiative transfer simulations using the dust distributions at $t=3\times10^{5}$ yr shown in Section \ref{sec:dustpileup}.
Table \ref{table:diskmodel} summarize the disk models that we used in the radiative transfer simulations.

\begin{table}[ht]
  \caption{Disk models used in the radiative transfer simulations}
  \label{table:diskmodel}
  \centering
  \begin{tabular}{cccc}
    \hline
    Model  & $\alpha_{\rm dead}$  &  $v_{\rm frag}~[{\rm m~s^{-1}}]$ & Note\\
    \hline \hline
    Model 1  & $10^{-3}$  & 1 & $0.01\Sigma_{\rm d}$\\
    Model 2  & $10^{-3}$  & 10 &\\
    Model 3  & $10^{-4}$  & 10 &\\
    Model 4  & $10^{-4}$  & 10 & Equation \eqref{eq:plts}\\
    \hline
  \end{tabular}
\end{table}

Model 1 ($\alpha_{\rm dead}=10^{-3}$ and $v_{\rm frag}=1~{\rm m~s^{-1}}$) has no dust-pileup, while the others have a dust-pileup at the dead-zone inner edge.
In model 2, 3 and 4, we directly use the dust distribution shown in Figure \ref{fig:sigma}, while we have a modification in model 1.
In model 1, the dust surface density is reduced by a factor of 100 from that obtained by the dust-growth simulation to avoid non-convergence (see Section \ref{sec:instability}). 
The effect of planetesimal formation (Equation \eqref{eq:plts}) is included only in model 4.

The radial coordinate ranges from $0.03~{\rm au}$ to $1000~{\rm au}$ and is logarithmically divided into 100 grids per one decade.
The theta coordinate (angle from z-axis) ranges from $60^{\circ}$ to $90^{\circ}$ and is linearly divided into 96 grids. We assume that the disk is axisymmetric.
For each radial bin, we assume a dust size distribution ranging from $0.1~{\rm \mu m}$ to $a_{\rm max}=(3m_{\rm p}/4\pi\rho_{\rm int})^{1/3}$ that follows a power law with a index of $-3.5$ similar to the MRN distribution \citep{MRN1977}.
In order to allow settling of different sizes of dust particles to a different height in the disk, the dust size distribution is divided into five groups, $<0.3$, 0.3--3, 3--30, 30--300, 300--3000 ${\rm \mu m}$, having a representative size of 0.1, 1, 10, 100, 1000 ${\rm \mu m}$, respectively.
We ignore the contribution from dust particles larger than $3000~{\rm \mu m}$ because it has little effect on the result within the range of wavelength we focus on.
For each dust size bin, the dust scale height is calculated by Equation \eqref{eq:dustheight}.

The dust opacities are computed using Mie theory.
For simplicity, dust particles are assumed to be a single species of amorphous silicate (${\rm Mg_{0.7}Fe_{0.3}SiO_{{3}}}$).
The optical constants are taken from the Jena database\footnote{\url{http://www.astro.uni-jena.de/Laboratory/OCDB/}} \citep{Jaeger1994, Dorschner1995}.

We assume isotropic scattering for simplicity and employ the modified random walk approximation \citep{Min2009} in order to treat a very optically thick region owing to the dust-pileup. 
The central star is treated as a finite-size sphere.
The radiative transfer simulations are iteratively performed to obtain a robust structure of the disk and we use $10^{8}$ photon packages for each simulation.
In Appendix \ref{sec:app2}, we show that the number of photon packages of $10^{8}$ is large enough to obtain a well converged temperature profile.
The vertical density structure was iterated based on the thermal structure obtained from previous radiative transfer calculation following the approach of \citet{Kama2009}.
The iteration is performed until the deviation of the midplane temperature in each iteration is less than 10\%.

\subsection{Results of radiative transfer Simulations}
\subsubsection{Radial intensity profile}\label{sec:radialprofile}

\begin{figure}[ht]
\begin{center}
\includegraphics[scale=0.5]{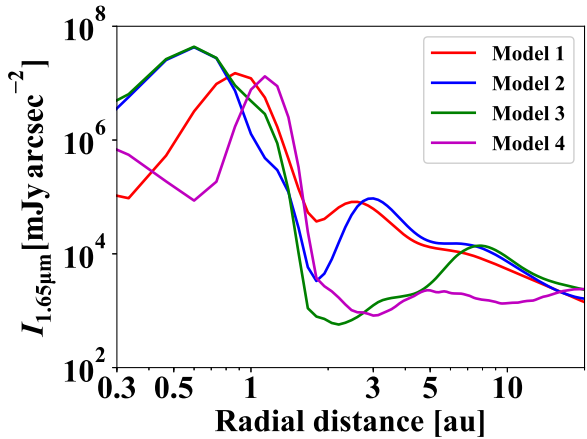}
\caption{Radial intensity profiles at $1.65~{\rm \mu m}$ for the disks with different values of $\alpha_{\rm dead}$ and $v_{\rm frag}$.
The intensity is the value at the major axis of the disk (along the long part of the ellipse) with inclination of $45^{\circ}$.
{\it red}: $\alpha_{\rm dead}=10^{-3}$ and $v_{\rm frag}=1~{\rm m~s^{-1}}$ with reducing the dust surface density by a factor of 100, 
{\it blue}: $\alpha_{\rm dead}=10^{-3}$ and $v_{\rm frag}=10~{\rm m~s^{-1}}$, 
{\it green}: $\alpha_{\rm dead}=10^{-4}$ and $v_{\rm frag}=10~{\rm m~s^{-1}}$,
{\it magenta}: $\alpha_{\rm dead}=10^{-4}$ and $v_{\rm frag}=10~{\rm m~s^{-1}}$ with planetesimal formation effect. 
}
\label{fig:scatteredlight}
\end{center}
\end{figure}

Figure \ref{fig:scatteredlight} shows the radial intensity profiles at $1.65 {\rm \mu m}$ (H-band) for the disks with different values of $\alpha_{\rm dead}$ and $v_{\rm frag}$.
We clearly see a large bump in the radial intensity profile around $0.9~{\rm au}$ in model 1, $0.6~{\rm au}$ in model 2 and 3 and $1.1 {\rm au}$ in model 4.
This bright ring is caused by the direct irradiation on the inner rim of the dust disk (see \citealt{Dullemond2010} for review).
In model 1, because we artificially reduced the dust surface density, the radial optical depth for the stellar light is smaller than that in model 2 and 3, and hence the dust rim occurs at a larger distance from the star. 
Model 4 also has a dust rim at a larger distance ($\sim 1.1 {\rm au}$).
This is because dust particles are converted into planetesimals at the dead-zone inner boundary before they leak out of the dead-zone inner boundary, resulting into the depletion of dust particles inside the dead-zone inner boundary.

Although there is a dust wall at the dead-zone inner boundary in model 2, 3 and 4, it is hard to distinguish it from the bright inner rim of dust disk because the separation between the inner rim and the dust wall is almost comparable to the width of the inner rim.
Especially in model 4, the dust wall at the dead-zone inner boundary coincides with the inner rim of the dust disk because of the depletion of dust particles inside the dead-zone inner boundary.

The most important difference in the radial intensity profile of the four models is in the deep intensity dip beyond $1~{\rm au}$.
This is an indication of a shadow mainly caused by the dust-pileup at the dead-zone inner edge, with small additional contribution from the puffed-up inner rim.
In model 3, the intensity decreases by five orders of magnitude around the dead-zone inner edge from inside out and the shadowed region extends to $\sim 8~{\rm au}$.
In this region, the angle between the disk surface and incident stellar light is nearly zero, meaning that disk surface does not receive direct irradiation from the central star (see Figure \ref{fig:2d}(a) in Appendix \ref{sec:app3}).
In model 4, because the dust wall acts even as a inner rim of the dust disk, the combined effect results in the shadow extending to $\sim 20~{\rm au}$.

The effect of the dust shadow is more significant for smaller $\alpha_{\rm dead}$ and larger $v_{\rm frag}$.
There are two reasons for this trend.
Firstly, smaller $\alpha_{\rm dead}$ and larger $v_{\rm frag}$ lead to more efficient trapping of dust particles as mentioned in Section \ref{sec:dustpileup}.
Secondly, smaller $\alpha_{\rm dead}$ and larger $v_{\rm frag}$ result in a smaller amount of small particles behind the dead-zone inner edge because the particles grow and drift inward without catastrophic fragmentation.
As explained below, this effect strongly depends on these parameters.
With the assumption that the surface number density of dust particles per unit particle size is proportional to $a^{-3.5}$, one can show that the surface number density in each dust-size bin is approximately proportional to $\Sigma_{\rm d}a_{\rm max}^{-1/2}$. The dust surface mass density is approximately inversely proportional to the maximum Stokes number, which is proportional to $a_{\rm max}$.
Therefore, the number of small particles behind the dead-zone inner edge is proportional to $a_{\rm max}^{-3/2}$.
If we combine this with Equation \eqref{eq:typicalst}, the number of small particles in the region just behind the dead-zone inner edge is proportional to $\alpha_{\rm dead}^{-3/2}v_{\rm frag}^{3}$, which indicates that the width and depth of the dust shadow is sensitive to these parameters.

\subsubsection{Synthetic images}
\begin{figure*}[ht]
\begin{center}
\includegraphics[scale=0.45]{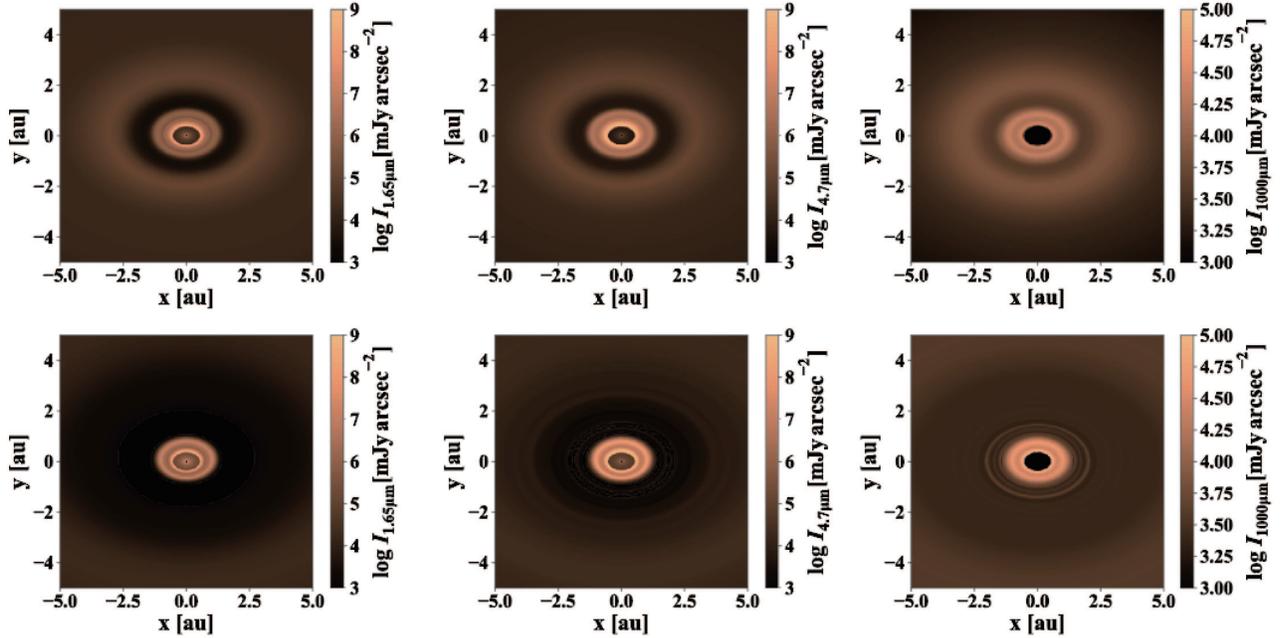}
\caption{Synthetic images of the inner region of disks for model 2 (top) and 3 (bottom) viewed $45^{\circ}$ from face-on. 
The intensity maps are at wavelengths 1.65 (left), 4.7 (center) and 1000 ${\rm \mu m}$ (right) in log-scale.
}
\label{fig:sca2d}
\end{center}
\end{figure*}
To provide a realistic view of the inner region of disks, we constructed synthetic images of our models.
Figure \ref{fig:sca2d} shows the images of the inner region of disks with model 2 and 3 for an inclination of $45^{\circ}$.
The top panels correspond to model 2 and the bottom panels to model 3.
The images are at wavelengths 1.65, 4.7 and $1000~{\rm \mu m}$ from left to right.
In model 2, we observe a sharp shadow with a width of $\sim 1~{\rm au}$ behind the dead-zone inner edge, while the shadow is less sharp in model 3 due to its large width ($\sim 9~{\rm au}$).
The dust halo in front of the inner rim of dust disk is brightest at wavelength $1.65~{\rm \mu m}$, while the dust rim itself is brighter at $4.7~{\rm \mu m}$ as described by \citet{Flock2016}.
At longer wavelengths, the width of shadow is smaller.
When seen at 1 mm, the brightness of the shadowed region differs only by a factor of $<5$ between models 2 and 3. At 1.65 ${\rm \mu m}$, in contrast, the brightness differs by two orders of magnitude.

\begin{figure}[ht]
\begin{center}
\includegraphics[scale=0.55]{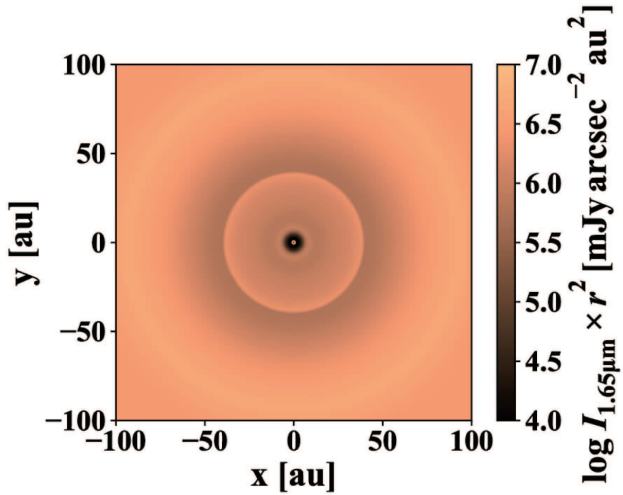}
\caption{
Intensity map of the disk model 3 with the field width of $200~{\rm au}$ at wavelength $1.65~{\rm \mu m}$ viewed from face-on.
The intensity is multiplied by $r^{2}$ to compensate for the stellar flux attenuation.
}
\label{fig:sca2d-100au}
\end{center}
\end{figure}
We also provide a intensity map of the disk model 3 at $1.65~{\rm \mu m}$ with the field of view of $200~{\rm au}$ wide (Figure \ref{fig:sca2d-100au}, the intensity is multiplied by $r^{2}$).
Figure \ref{fig:sca2d-100au} shows that the intensity distribution has a central hole of 10 au in radius, corresponding to the shadow casted by the dust-pileup.
In addition to that, we observe another shadow extending from $~40~{\rm au}$ outward.
This extended outer shadow is caused by the water snow line (see also \citealt{Pinilla2017}).
Outside the snow line, dust particles can grow large and settle onto the midplane, while dust particles just inside the snow line are smaller due to ice sublimation and fragmentation.
The abrupt change in the size of dust particles around the snow line creates a shadow behind it.

\subsubsection{Midplane temperature} \label{sec:temperature}
\begin{figure}[ht]
\begin{center}
\includegraphics[scale=0.5]{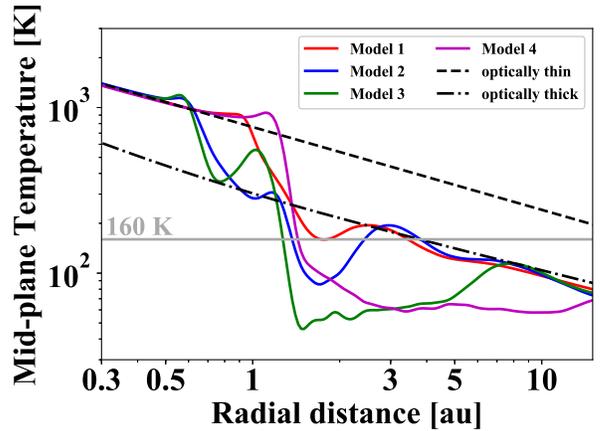}
\caption{Midplane temperature profile of the disks with different values of $\alpha_{\rm dead}$ and $v_{\rm frag}$. 
{\it red}: $\alpha_{\rm dead}=10^{-3}$ and $v_{\rm frag}=1~{\rm m~s^{-1}}$, 
{\it blue}: $\alpha_{\rm dead}=10^{-3}$ and $v_{\rm frag}=10~{\rm m~s^{-1}}$, 
{\it green}: $\alpha_{\rm dead}=10^{-4}$ and $v_{\rm frag}=10~{\rm m~s^{-1}}$,
{\it magenta}: $\alpha_{\rm dead}=10^{-4}$ and $v_{\rm frag}=10~{\rm m~s^{-1}}$ with planetesimal formation. 
Black dashed and dot-dashed lines denote the temperature profile given by Equation \eqref{eq:temp} and that of the optically thick disk (Equations (11)-(15) in \citealt{Ueda2017} with $z_{\rm *,D}=0.5h_{\rm g}$).
Gray horizontal line corresponds to the sublimation temperature of water ice ($T=160~{\rm K}$).
}
\label{fig:tempall}
\end{center}
\end{figure}

In Figure \ref{fig:tempall}, we show the midplane temperature derived from the thermal Monte Carlo simulations done by RADMC-3D \citep{RADMC}.
In model 1, an optically thin inner disk connects smoothly to an optically thick outer disk, with a shallow dip in the temperature profile at $\sim 2~{\rm au}$ caused by a shadow. 
As mentioned above, there is no dust-pileup in model 1, so this shadow simply originates from the puffed-up inner rim \citep{DDN2001, Flock2016}.
In model 2, 3 and 4, we observe a wide dip in the midplane temperature behind $\sim 1~{\rm au}$.
Especially for model 3 ($\alpha_{\rm dead}=10^{-4}$ and $v_{\rm frag}=10~{\rm m~s^{-1}}$), the temperature in the shadowed region has a minimum of as low as 40 K.
In Appendix \ref{sec:app3}, we examine the effect of scattering by switching off it and find that the minimum temperature in the shadowed region is insensitive to scattering.

Due to the shadowing effect, the location of the water snow line is completely different from what is expected from the temperature profiles of smooth disks.
Assuming the water sublimation temperature of 160 K, the water snow line in the smooth, optically thin (with an assumption that grains behave as a black body) and thick disks is located at $\approx 20$ au and $\approx 3$ au, respectively. 
In comparison, the water snow line in model 3 and 4 lies at $\approx 1.3~{\rm au}$, which is just behind the dead-zone inner edge.
In model 2, there are three water snow lines located respectively at $\approx 1.3$, $2$ and $4~{\rm au}$ because of the local temperature dip lying at 1--2 au.
These multiple snow lines would affect the evolution of dust particles through the sublimation and recondensation of water ice as we discuss in Section \ref{sec:diss}.

\section{Waves on the disk surface}\label{sec:instability}
In this section, we show the results of the radiative transfer simulations of model 1 without the reducing factor for the dust surface density (in other words, we use the original dust distribution obtained from the dust-growth simulation).
In this model, we observe oscillatory behavior in the temperature structure probably connected to the so-called thermal wave instability \citep{DAlessio1999, Dullemond2000, Watanabe2008}.
For this reason, the temperature profile does not relax into a steady state (see more details in Appendix \ref{sec:app}).

Figure \ref{fig:iteration-1period} illustrates the oscillatory behavior of the temperature profile.
We here show the midplane temperature at the 17th to 30th iteration steps.
We see that the temperature structure is stable in the innermost region ($< 1~{\rm au}$) and in the outermost region ($>100~{\rm au}$), but oscillates in the intermediate region ($1~{\rm au}<r<100~{\rm au}$).
\begin{figure}[ht]
\begin{center}
\includegraphics[scale=0.5]{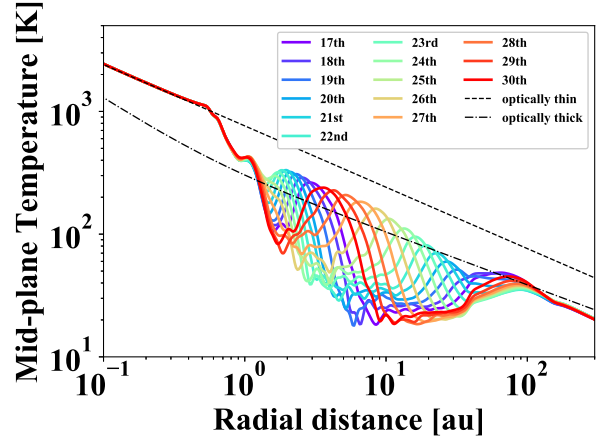}
\caption{Midplane temperature of model 1 without the reducing factor for the dust surface density at the 17th to 30th iteration steps. 
Black dashed and dot-dashed lines are the same as in Figure \ref{fig:tempall}.
}
\label{fig:iteration-1period}
\end{center}
\end{figure}
The oscillatory behavior is caused by waves propagating inward along the disk surface as reported by \citet{Watanabe2008} and \citet{Min2009}.
Figure \ref{fig:sca1period} shows the intensity profiles at wavelength $1.65~{\rm \mu m}$ at the 17th to 30th iteration steps.
\begin{figure}[ht]
\begin{center}
\includegraphics[scale=0.5]{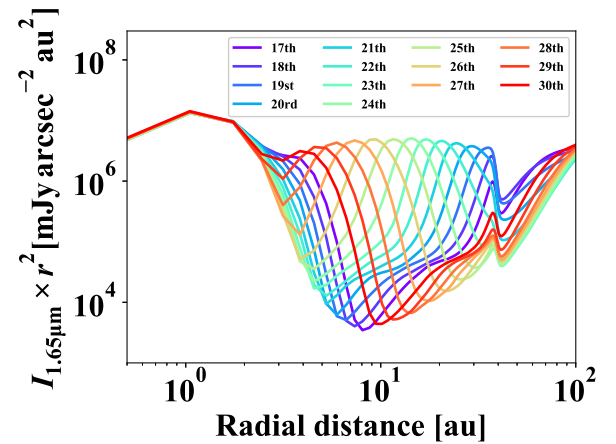}
\caption{
Radial intensity profile of model 1 without the reducing factor for the dust surface density at the 17th to 30th iteration steps.
The intensity is the value at the major axis of the disk with inclination of $45^{\circ}$.
The intensity is multiplied by $r^{2}$ to compensate for the stellar flux.
}
\label{fig:sca1period}
\end{center}
\end{figure}
We see that the shadows and bright ring-like structures in the radial intensity profile move inward and decay around $2~{\rm au}$.
Once a bump is created on the surface by the perturbation, the illuminated front side of the bump receives more stellar light and the shady back side of the bump receives insufficient flux.
At the illuminated side, the disk surface puffs up further as the midplane temperature increases, resulting in further decrease in the temperature at the back side.
The oscillatory behavior is not related to the shadow casted by the water snow line ($\sim 40~{\rm au}$, see also Figure \ref{fig:sca2d-100au}).
We confirmed that the oscillatory behavior still exists even without the water snow line.

The reason why only model 1 shows the oscillatory behavior is not clear, but it might be linked with the optical thickness.
Model 1 has the highest optical depth in our models because of efficient fragmentation of dust particles.
We found that in model 1, the amplitude of the oscillation decreases as the dust surface density is reduced.
And also we confirmed that even in the other models, the oscillatory behavior is seen if we artificially increase the dust surface density by a factor of 30--100.

We remind that our radiative transfer simulations assume hydrostatic equilibrium of vertical structure and calculate equilibrium temperature structure for a given dust distribution at each iteration step, meaning that our approach does not treat the time evolution of the instability and does not necessarily give us the realistic behavior of it.
In order to know the realistic behavior of the instability, it is important to calculate the vertical heat diffusion and hydro dynamics time-dependently because the height of the disk surface where the stellar irradiation is received depends on the midplane temperature.

\section{Discussion} \label{sec:diss}

\subsection{Implications for planet formation}
We demonstrated that the total mass of planetesimals formed at the dead-zone inner edge strongly depends on the turbulence strength and the critical fragmentation velocity. 
As we mentioned in Section \ref{sec:dustpileup}, the total planetesimal mass is $1M_{\oplus}$ for disk with $\alpha_{\rm dead}=10^{-3}$ and $v_{\rm frag}=10~{\rm m~s^{-1}}$, while it is $705M_{\oplus}$ for disk with $\alpha_{\rm dead}=10^{-4}$ and $v_{\rm frag}=10~{\rm m~s^{-1}}$.
The large difference in the total planetesimal mass caused by the difference in the turbulence strength might link with the difference in the formation of the solar system and super-earth systems.
Our results suggest that systems with low-mass terrestrial planets such as the solar system would form in the disk with relatively strong turbulence ($\alpha_{\rm dead}\gtrsim10^{-4}$), while systems with super-earths would form in the disk with weak turbulence ($\alpha_{\rm dead}\lesssim10^{-4}$).
Recent 3D MHD simulations suggest that the turbulence strength in the dead-zone is weaker than $10^{-4}$ (e.g., \citealt{Gressel2015}; \citealt{Flock2017}), indicating that super-earth systems should be more common than systems like the solar system.
Although the dead-zone inner edge lies much closer to the central star ($\sim 0.1~{\rm au}$) for T-Tauri disks, \citet{Ogihara2018} showed that outward migration of planetesimals induced by the disk wind may have led to the formation of terrestrial planets in the solar system.

The non-monotonic radial temperature profile caused by the shadow would affect the water mass fraction of inner planets.
\citet{Grimm2018} showed that in the TRAPPIST-1 system, planet d has a relatively low density ($\sim 3.4~{\rm g~cm^{-3}}$) while the other outer planets has higher density ($\sim$ 4--5.6 ${\rm g~cm^{-3}}$).
This density distribution suggests that planet d might have a higher water mass fraction ($>5\%$) than that of the other outer planets \citep{Grimm2018}.
The inner icy region caused by the shadow might explain the origin of water in the inner planet.

\subsection{Feedback from the shadows on dust growth}
As shown in Section \ref{sec:temperature}, the shadows affect the temperature structure and the resultant temperature differs considerably from that used in the dust growth simulations.
In particular, in models 2, the shadow causes a local dip of the midplane temperature below the ice-sublimation temperature,
producing two snow line in addition to the one lying at $\approx 4~\rm au$ that would be present without the shadow.
The multiple water snow lines would result into the multiple bright and dark rings in millimeter and infrared observations as proposed by \citet{Pinilla2017}.
If icy dust particles recondense in the shaded region, the particles can grow larger and settle down to the disk midplane, which would enhance the shadowing effect by lowering the local disk surface.

The thermal wave instability would also affect the evolution of dust particles.
As shown in Section \ref{sec:instability}, the thermal waves create the positive temperature gradient in front of the peak.
If the pressure gradient is positive, dust particles drift outward.
In our simulations, the sign of pressure gradient does not be positive but the absolute value is less than unity (see Appendix \ref{sec:app}), indicating that the thermal waves might help dust particles to grow larger beyond the radial drift barrier.
Investigation of these possibilities requires dust growth simulations coupled with radiation hydrodynamical calculations, which will be the subject of future work.

\subsection{Implications for disk observations}
As shown in Figures \ref{fig:scatteredlight}, it is not easy to directly detect a bright ring structure originating from the dust-pileup at the dead-zone inner edge because it is too close to the inner rim.
However, the shadow casted by the dust-pileup would potentially appear in the observations. 
As shown in Figures \ref{fig:scatteredlight} and \ref{fig:sca2d-100au}, the shadow extends to $\sim 10~{\rm au}$ and significantly reduces the temperature at the shaded region.
As a result, the mid-infrared emission which usually emitted from the region where temperature is 300--1000 K is significantly reduced.
In model 4 of our simulation, the MIR disk ``size'', the region where the half of the total mid-infrared emission from the system is released within it, is almost the same as that in near-infrared wavelength.
It means that the most of the mid-infrared emission comes from the inner rim as near-infrared emission does so.
Although \citet{MillanGabet2016} pointed out that the observed MIR disk sizes is much smaller than that expected from the standard flared disk model, they also showed that it is still larger than that in near-infrared wavelengths. 
Future high resolution observations at (mid-)infrared wavelength using such as VLTI/MATISSE, E-ELT and TMT will allow us to directly compare our models with the observations.
For example, E-ELT/METIS (e.g., \citealt{Brandl2014}) would have an angular resolution of $5~{\rm mas}$ at a wavelength of $10~{\rm \mu m}$, which provides us images of the inner region of disks at a distance of $100~{\rm pc}$ with the spatial resolution of $0.5~{\rm au}$.

The thermal wave instability also creates shadows and ring-like structures on the disk surface.
Recent infrared observations have revealed that many circumstellar disks have ring-like structures on its surface (e.g., \citealt{Avenhaus2018}; \citealt{Bertrang2018}).
Although such structures are sometimes interpreted as an evidence of on going formation of giant planets, the thermal wave instability would be another explanation.
The disk around TW Hya would be a nice example having three bright ring-like structures \citep{Boekel2017, Akiyama2015} possibly caused by the thermal wave instability.
One possible way to distinguish the origins of these rings is to measure the gas surface density profile by using molecular line emission (e.g., ${\rm C^{18}O}$, \citealt{Nomura2016}).
The gas surface density must have gaps if the shadows and rings are caused by planets, while the thermal wave instability does not create large gaps in the gas surface density profile since it is associated with the change in the temperature structure.

\section{Summary} \label{sec:sum}
We performed simulations of one-dimensional dust and gas disk evolution with fully including the backreaction from dust to gas to investigate the observational features of a dust-pileup at the dead-zone inner edge.
Around the dead-zone inner edge, dust particles are easy to fragment into small particles because of high speed collisions due to the turbulent motion, which interferes with the dust-pileup.
We demonstrated that a strong dust pile-up at the dead-zone inner edge occurs only for condition where the Stokes number of dust particles is much larger than the turbulence strength in the dead-zone.
Based on the fact that the Stokes number at the dead-zone inner edge is determined by the turbulence induced fragmentation, we derived the condition for the dust-pileup as a function of the critical fragmentation velocity of silicate dust particles and the turbulence strength, and found that if the critical fragmentation velocity is $1~{\rm m~s^{-1}}$, the turbulence parameter in the dead-zone needs to be lower than $3 \times 10^{-4}$ for dust trapping to operate.
The total mass of planetesimals formed at the dead-zone inner boundary is quite sensitive to the turbulence strength, which might explain the diversity in the mass of the inner planets such as super-earths and terrestrial planets in our solar system.

Using the dust distribution obtained above, we also performed the radiative transfer simulations with RADMC-3D to construct models of the inner region of protoplanetary disks including the effect of the dust-pileup.
We found that if dust particles strongly concentrate at the dead-zone inner edge, the dust-pileup acts as an optically thick wall, casting a $10~{\rm au}$-scale shadow directly behind the dead-zone inner edge.
The shadow significantly reduces the midplane temperature.
The resultant temperature profile suggests that the water snow line could be much closer to the central star and potentially creates multiple water snow lines.

Even if dust particles do not pile up on the dead-zone inner edge owing to efficient fragmentation, waves are naturally excited on the disk surface, possibly by the so-called thermal wave instability, creating shadows and ring-like structures on the disk surface.
These ring-like structures might account for the bright rings seen in the scattered light images of some disks, including that of TW Hya \citep{Boekel2017}.
The waves create a positive temperature gradient in front of the peak, which might help dust particles to grow larger beyond the radial drift barrier.

Future high resolution observations at (mid-)infrared wavelength using such as VLTI/MATISSE, E-ELT and TMT will allow us to directly compare our models with the observations.
\acknowledgments
We would like to thank C. P. Dullemond, for useful comments.
This work was supported by JSPS KAKENHI Grant Numbers JP18J14595, JP16K17661.
M.~F. has received funding from the European Research Council (ERC) under the European Union's Horizon 2020 research and innovation programme (grant agreement No. 757957).
\software{RADMC-3D \citep{RADMC}}



\appendix
\section{Details of the oscillatory behavior}\label{sec:app}
In this appendix, we provide the details of the oscillatory behavior shown in Section \ref{sec:instability}.
\begin{figure}[ht]
\begin{center}
\includegraphics[scale=0.5]{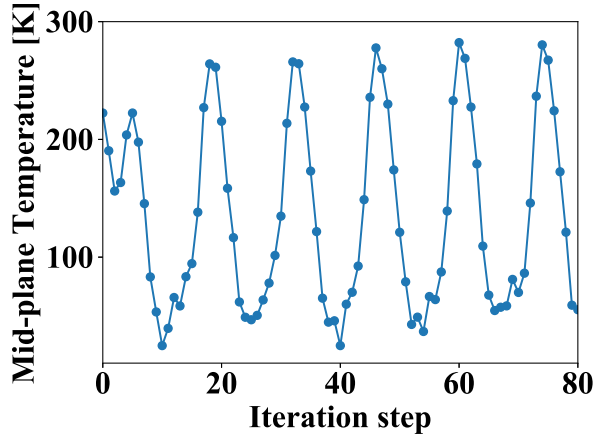}
\caption{Midplane temperature at $3~{\rm au}$ of model 1 without the reducing factor for the dust surface density at each iteration step.
}
\label{fig:iteration-3au}
\end{center}
\end{figure}
Figure \ref{fig:iteration-3au} shows the midplane temperature at $3~{\rm au}$, which is within the unstable region.
The temperature oscillates around $150~{\rm K}$ with an amplitude of $130~{\rm K}$ with a period of 14 iterations.
We confirmed that the periodic oscillation does not converge at least within 80 iterations.
\begin{figure}[ht]
\begin{center}
\includegraphics[scale=0.5]{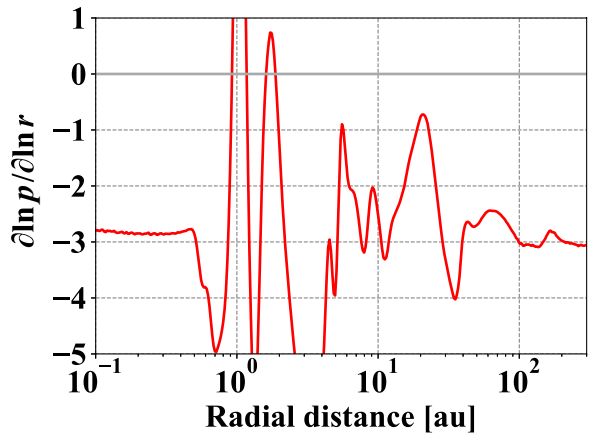}
\caption{
Logarithmic pressure gradient $\partial \ln p/ \partial \ln r$ in the simulation after 20 iterations.
The horizontal gray solid line denotes $\partial \ln p/ \partial \ln r=0$ where the direction of the dust radial motion is reversed.
}
\label{fig:gradp}
\end{center}
\end{figure}
Figure \ref{fig:gradp} shows the logarithmic pressure gradient $\partial \ln p/ \partial \ln r$ in the simulation after 20 iterations.
The local pressure maximum around $1~{\rm au}$ is associated with the dead-zone inner edge.
In addition to that, there are two peaks in $\partial \ln p/ \partial \ln r$ around $\sim 2~{\rm au}$ and $\sim 20~{\rm au}$ respectively.
These peaks correspond to the illuminated side of the waves on the disk surface.
The intense irradiation onto the illuminated side of the waves creates the positive temperature gradient which makes the disk gas rotates with Keplerian frequency.
When the pressure gradient is negative/positive, dust particles drift inward/outward.
Although the value of $\partial \ln p/ \partial \ln r$ at these peaks is not positive, the absolute value is 4--6 times smaller than the fiducial value ($\sim 3$).

\section{Convergence of the Thermal Monte Carlo simulation}\label{sec:app2}
We used $1\times 10^{8}$ photon packages for each thermal Monte Carlo simulation.
In order to investigate if the thermal Monte Carlo simulations converge well, we additionally performed simulations with $2 \times 10^{9}$ photon packages, which is 20 time higher than fiducial.
\begin{figure}[ht]
\begin{center}
\includegraphics[scale=0.5]{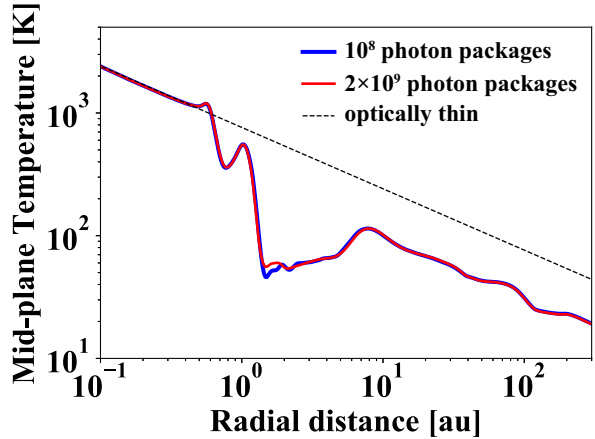}
\caption{
Midplane temperature for disk model 3 obtained from the thermal Monte Carlo simulations using $1\times 10^{8}$ (blue) and $2\times 10^{9}$ (red) photon packages.
The black dashed line denotes the temperature profile given by Equation (\ref{eq:temp}).
}
\label{fig:convergence-test}
\end{center}
\end{figure}
Figure \ref{fig:convergence-test} shows the midplane temperature profile for disk model 3 obtained from the thermal Monte Carlo simulations for the two different numbers of photon packages.
We find no significant difference between the two simulation results, indicating convergence. 
We also confirmed that the number of photon packages of $1\times 10^{8}$ is large enough to obtain a well converged temperature profile for the other disk models.

\section{Effect of Scattering}\label{sec:app3}
In this section, we examine how dust scattering affects the temperature structure of the disk.
\begin{figure}[ht]
\begin{center}
\includegraphics[scale=0.5]{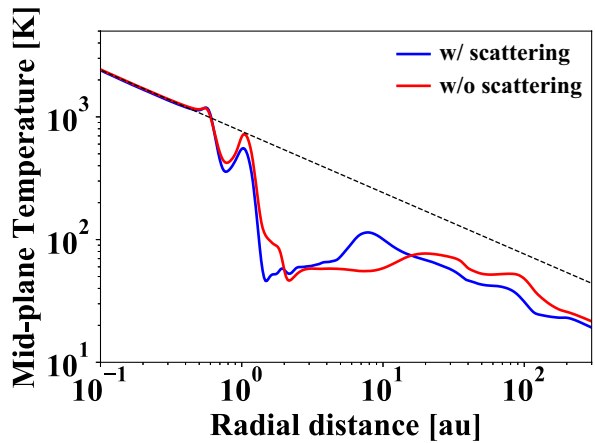}
\caption{
Midplane temperature profile for disk model 3 obtained from radiative transfer simulations with (blue) and without (red) scattering.
The black dashed line denotes the temperature profile given by Equation (\ref{eq:temp}).
}
\label{fig:scattering}
\end{center}
\end{figure}
Even if the disk cannot receive the direct irradiation from the star, photons scattered from the upper layer could heat the disk interior.
In order to investigate the effect of scattering, we performed radiative transfer simulations with scattering switched off.
Figure \ref{fig:scattering} shows the results with and without scattering for model 3.
We find that the radial width of the shadow is narrower when scattering is switched on but the minimum midplane temperature in the shadowed region is insensitive to scattering.
\begin{figure}[ht]
\begin{center}
\includegraphics[scale=0.42]{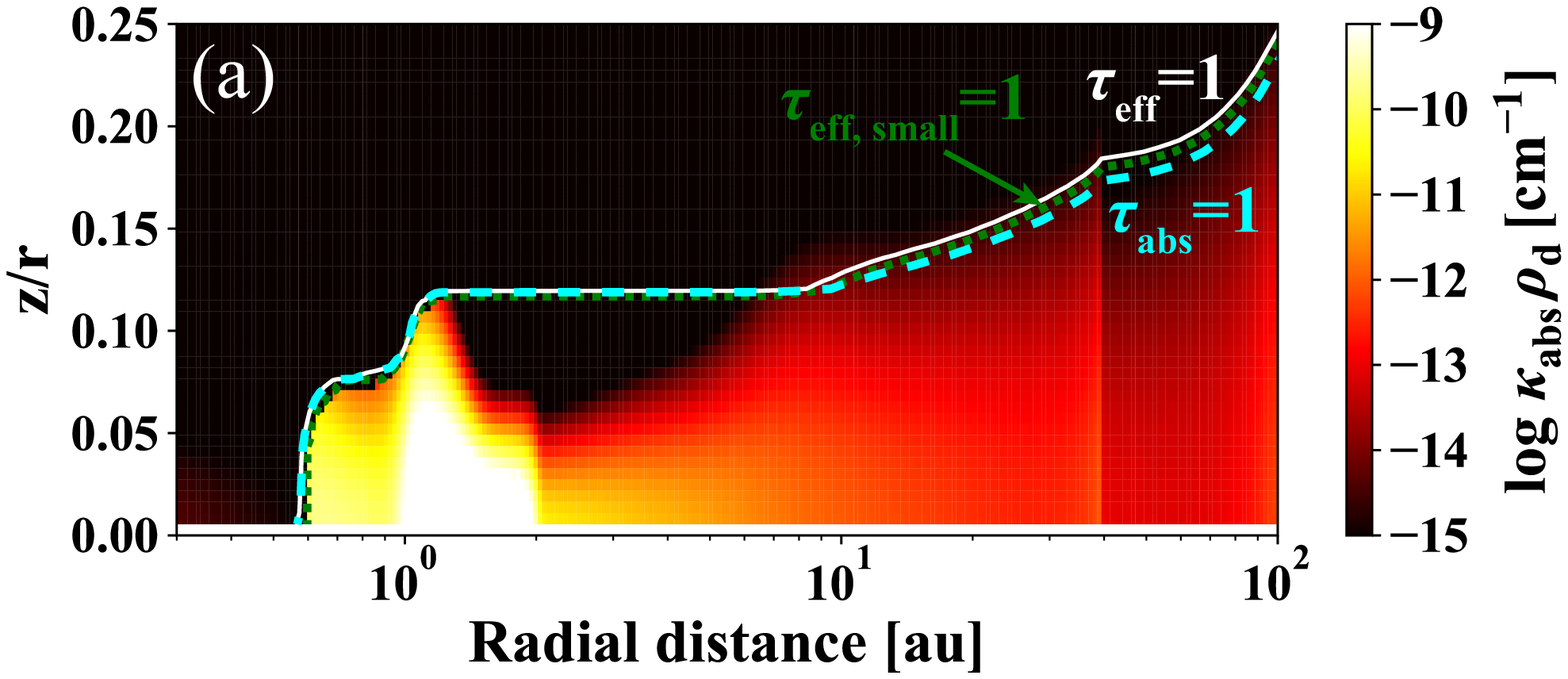}\includegraphics[scale=0.42]{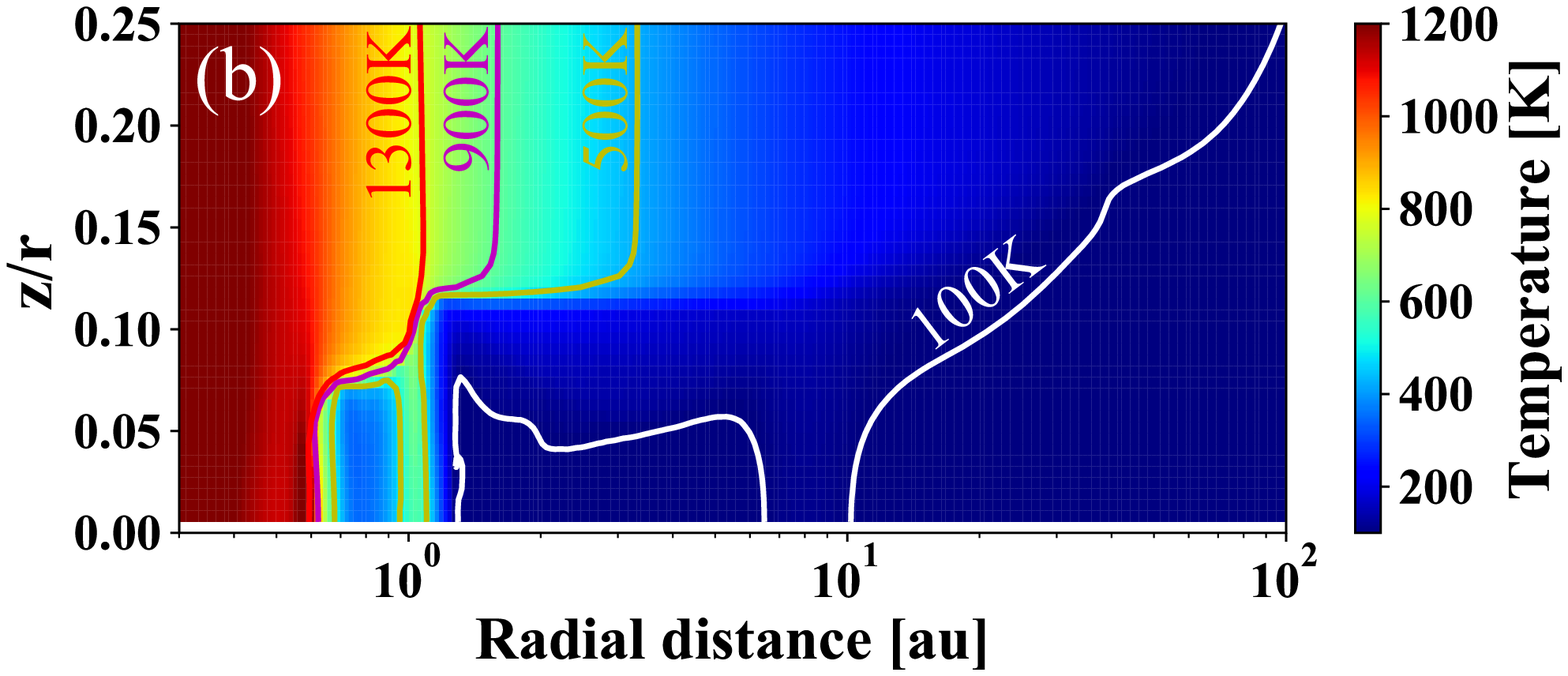}
\caption{
Two-dimensional structure of disk model 3.
(a) Distribution of the dust absorption opacity $\kappa_{\rm abs}\rho_{d}$ for the wavelength of $0.3~{\rm \mu m}$ contributed from the smallest dust-size bin ($<0.3~{\rm \mu m}$). 
The white solid line denotes the effective absorption surface where
the total effective optical depth $\tau_{\rm eff}$ is equal to unity.
The cyan dashed line denotes the absorption surface where the absorption optical depth $\tau_{\rm abs}$ is equal to unity.
The green dash-dotted line denotes the location where the effective optical depth contributed from the smallest dust-size bin $\tau_{\rm eff, small}$ is equal to unity.
(b) Temperature distribution with contour lines of 100K (white), 500K (yellow), 900K (purple) and 1300K (red).
}
\label{fig:2d}
\end{center}
\end{figure}

Figure \ref{fig:2d}(a) shows the two-dimensional distribution of the dust absorption opacity $\kappa_{\rm abs}\rho_{\rm d}$ for the wavelength of $0.3~{\rm \mu m}$ contributed from the smallest dust-size bin.
We also plot the effective absorption surface where the effective optical depth $\tau_{\rm eff}$ is equal to unity and the absorption surface where the absorption optical depth $\tau_{\rm abs}$ is equal to unity.
The effective absorption optical depth is calculated using the effective absorption coefficient $\kappa_{\rm eff}=\sqrt{\kappa_{\rm abs}(\kappa_{\rm abs}+\kappa_{\rm sca})}$ \citep{Radipro} where $\kappa_{\rm abs}$ and $\kappa_{\rm sca}$ are respectively the absorption and scattering coefficient.
The effective absorption coefficient is useful to express how much distance a photon travels before it is truly absorbed by the medium with taking into account of the effect of scattering.
We see that the effective absorption surface is not so different from the absorption surface, which means that scattering does not change the surface structure so much.
Figure \ref{fig:2d}(a) also shows that the effective optical depth is almost determined by the smallest grains.
The height of absorption surface in the $r$-$z/r$ plane is constant between 1 and $8~{\rm au}$,
In this region, the disk surface does not receive direct irradiation from the central star.
Owing to the shadowing effect, the midplane temperature in the shadowed region significantly decreases as shown in Figure \ref{fig:2d}(b).




\bibliographystyle{./aasjournal}
\bibliography{./DIB}



\end{document}